\documentclass{article}
\usepackage{arxiv}

\RequirePackage{amsthm,amsmath}
\RequirePackage[colorlinks,citecolor=blue,urlcolor=blue]{hyperref}

\RequirePackage{enumitem}
\RequirePackage{algorithm}
\RequirePackage{graphicx}
\RequirePackage{tikz}
\usetikzlibrary{bayesnet}
\usepackage[utf8]{inputenc} 
\usepackage[T1]{fontenc}    
\usepackage{url}            
\usepackage{booktabs}       
\usepackage{amsfonts}       
\usepackage{microtype}      
\usepackage{lipsum}		

\usepackage{graphicx}
\usepackage{doi}
\newcommand{\expect}{\text{E}}
\newcommand{\p}{p_{ik}^{(mn)}}

\usepackage[symbol]{footmisc}

\usepackage[style=authoryear,natbib, maxcitenames=2]{biblatex}
\title{Scalable Community Detection in Massive Networks using Aggregated Relational Data}

\author{Timothy Jones \\
	Department of Statistics\\
	Columbia University\\
	New York, NY 10027 \\
	\texttt{tdj2113@columbia.edu} \\
	\And
    Owen G. Ward \\
	Dept. of Statistics and Actuarial Science\\
	Simon Fraser University\\
	Burnaby, British Columbia \\
	\texttt{owen\_ward@sfu.ca} \\
	\And
	Yiran Jiang \\
	Department of Statistics\\
	Purdue University\\
	West Lafayette, IN 47907 \\
	\texttt{jiang693@purdue.edu} \\
	\And
    John Paisley \\
    Department of Electrical Engineering\\
	Columbia University\\
	New York, NY 10027 \\
	\texttt{jwp2128@columbia.edu} \\
	\And
	Tian Zheng\thanks {Corresponding author. Email: tian.zheng@columbia.edu} \\
	Department of Statistics\\
	Columbia University\\
	New York, NY 10027 \\
	\texttt{tian.zheng@columbia.edu} \\
}

\definecolor{red}{rgb}{0,0,0}
\newcommand{\supplementsection}{%
  \renewcommand{\thesection}{S\arabic{section}}
}

\begin{document}







\maketitle
\begin{abstract}
    The mixed membership stochastic blockmodel (MMSB) is a popular Bayesian network model for community detection. Fitting such large Bayesian network models quickly becomes computationally infeasible 
	when the number of nodes grows into hundreds of thousands and millions. 
    In this paper we propose a novel mini-batch strategy based on aggregated relational 
    data that leverages nodal information to fit MMSB to massive networks. 
	We describe a scalable inference method that can utilise nodal information 
	that often accompanies real-world networks. Conditioning on this extra information 
	leads to a model that admits a parallel stochastic variational inference algorithm, utilising stochastic gradients of 
    bipartite graph formed 
    from aggregated network ties between node subpopulations. 
	We apply our method to a citation network with over two million nodes and 25 
	million edges, capturing explainable structure in this network.
    Our method recovers parameters and achieves better convergence 
	on simulated networks generated according to the MMSB.
\end{abstract}



\section{Introduction}

Relational data between objects is commonly represented by a graph or network that encodes pairwise interactions,
and has been studied across the natural and social sciences. 
Due to this
prevalence in modern applications, analysis of such data is vitally important.
Among the tasks commonly
considered for such data,
community detection stands out as being one of the most crucial for practitioners and has
been widely studied.
Community detection algorithms aim to identify groups of nodes that exhibit similar connective behaviors. 
More specifically, nodes in networks often cluster into small communities, 
where nodes within a community 
{\color{red}show a similar propensity to form ties with other nodes}
\citep{bickel2009nonparametric}.
These clusters are often assortative, where nodes within a community interact more than nodes in 
different communities \citep{fortunato2016community}.
Identifying such clusters of nodes often provides important scientific insights 
into the processes underlying the realised network, and can also be used for further analysis
such as link prediction and node classification \citep{soundarajan2012using, ward2021next}.


The mixed membership stochastic blockmodel (MMSB) is a popular model-based method for 
community detection in networks\citep{airoldi2008mixed}. 
An extension of the widely used stochastic block model \citep{holland1983stochastic, nowicki2001estimation}, 
the MMSB relaxes the assumption that each node 
belongs to exactly one community, instead allowing the node to belong to different communities
for different interactions.
By allowing nodes to belong to multiple communities in varying amounts,
this model can 
capture more realistic features of networks \citep{airoldi2008mixed}.

A fundamental difficulty in analyzing networks with models such as the MMSB is the 
computational burden of fitting these flexible models to large real-world networks. 
The number of parameters grows quadratically in the number of nodes in the network. 
To mitigate this issue, many algorithms take advantage of the sparseness 
found in real-world networks by avoiding computations using all node pairs.
In particular, 
\citet{gopalan2013efficient} developed a stochastic variational inference (SVI) algorithm with a 
sub-sampling 
scheme that only uses links.

The computational burden of fitting network models may be reduced by including 
nodal information. Although the MMSB models connectivity only, 
real-world networks often have rich nodal data that can help with model convergence. Additionally, 
modeling nodal covariates reveal an interesting interplay between ``content" and ``connections" in networks. 
For instance, \citet{tan2016topic} modeled both text and links in a citation network to measure the topic-neutral impact of scientific articles.

Previous procedures for the MMSB have performed inference using the
fully observed adjacency matrix, including sampling links to scale this to 
large networks. These methods aim to infer the global network structure 
by sampling edges from the network in a possibly online fashion,
learning the parameters using gradient steps based on these small samples.
{\color{red} 
One obstacle to the use of} these methods is the inherent difficulty in sampling 
from networks. Such sampling schemes may only provide limited information
about a subset of the parameters in the model. For example for 
models describing community structure, it may be challenging to generate
a sample of nodes containing all possible pairs of community interactions.
This can lead to 
{\color{red} difficulties} in learning stable estimates and such models
may not converge efficiently.

In this paper we propose a novel mini-batch strategy based on aggregated relational 
data that leverages nodal information to fit MMSB to massive networks. 
Instead of forming a subgraph by sampling nodes or links, we create counts of 
links called aggregated relational data (ARD) for selected nodes in a mini-batch. 
These represent aggregate counts of interactions between a node 
and a specified subpopulation, derived from available nodal information.
These node subpopulation counts 
can be more readily analyzed than the original full adjacency matrix. These 
subpopulations could be formed 
utilising nodal information (such as the journals academic papers 
appear in).
The weighted bipartite graph formed by ARD retains more information from the original graph and 
can be used to estimate the parameters of an MMSB directly using the ARD data enumerated from the 
sampled mini-batches. We apply Bayesian modeling and derive a variational approximation algorithm 
for ARD that is used for each mini-batch. We show our strategy enjoys good computational 
efficiency and recovers true community structure in simulation experiments. A real-data analysis 
reveals insightful and meaningful structure 
from a citation network of scientific publications. 

This paper is organized as follows. 
Section~\ref{sec:background} provides a review of the MMSB model for network data, previous inference schemes for MMSB, and ARD models for network data. 
Section~\ref{sec:model} develops
our proposed extension, considering aggregate relational data for the mixed membership 
stochastic block model (ARDMMSB). In particular,
Section~\ref{sec:inference} introduces a new variational algorithm for 
approximate posterior inference for this model, describing important practical considerations
of our implementation. 
Section~\ref{sec:simulation} evaluates the performance of our proposed method
using simulated data while in Section~\ref{sec:empirical} we examine the performance of our proposed
for a large citation network.
In Section~\ref{sec:conclusion} we summarize our contributions and describe potential future applications of our model.

\section{Background}
\label{sec:background}

In this section, we review the Mixed-Membership Stochastic Blockmodel (MMSB) and past approaches for posterior inference for this model,
before providing a review of the idea of ARD.

\subsection{Review of MMSB}

\begin{figure}[t!]
\hrule
\vspace{0.2em}
\textbf{Mixed membership stochastic blockmodel (MMSB)}
\hrule
\begin{enumerate}
	\item For each entry of the blockmatrix $B$, draw probability $B_{mn} \sim \text{Beta}(a_{mn}, b_{mn}).$
	\item For each node $n = 1, \dots, N$, draw mixed membership vector $\pi_n \sim \text{Dirichlet} (\alpha).$
	\item For each node pair $(i, j)$, draw
	\begin{itemize}[itemsep=-1.5pt,leftmargin=*, topsep = 0pt]
		\item the sender membership indicator $s_{ij} \sim \text{Multi} (1, \pi_i)$ 
		\item the receiver membership indicator $r_{ij} \sim \text{Multi} (1, \pi_j).$ 
		\item the resulting interaction $\delta_{ij}\sim \text{Bernoulli} (B_{s_{ij} r_{ij}} ).$
	\end{itemize}
\end{enumerate}
\hrule
\caption{Data Generating Processes for the MMSB}
\label{fig:dgs_mmsb}
\end{figure}

The MMSB is a mixed membership latent variable
model for a directed graph that detects overlapping communities within a network. It assumes 
there 
are {\color{red}$K$} latent communities (groups). The global connectivity is stored in a block matrix 
$B$, where {\color{red} $B_{lm}$} represents the probability of a directed link from a node that assumes 
membership in group {\color{red}$l$}
to a node in group {\color{red}$m$}.
Each node belongs to the {\color{red}$K$}
groups with varying degrees of affiliation encoded in a {\color{red}$K$-dimensional} probability vector $\pi_i$. The membership vectors are drawn from a Dirichlet$(\alpha)$ prior. Each node $i$ may assume different memberships when interacting with different nodes. For each $(i,j)$, membership indicator vectors for the sender ($s_{ij}$) and receiver ($r_{ij}$) are drawn from a multinomial based on their membership vectors. If the {\color{red}$l$th}
and {\color{red}$m$th} elements of $s_{ij}$ and $r_{ij}$ are ones respectively, the value of the interaction $\delta_{ij}$ is sampled 
from 
Bernoulli({\color{red}$B_{lm}$}). Figure~\ref{fig:dgs_mmsb} outlines this data generating process.

\subsection{Review of prior posterior inference approaches for MMSB}
Since the MMSB models each directed edge, it quickly becomes computationally infeasible as the number of nodes grows. A popular approach to fit large Bayesian models is stochastic variational inference \citep{hoffman2013stochastic}. Variational approximation algorithms turn a Bayesian inference problem into an optimization procedure. The researcher first posits a family of approximating densities and the algorithm seeks to find the density that minimizes the Kullback-Leiber divergence to the exact posterior \citep{blei2017variational}. Stochastic variational inference implements gradient based optimization that combines natural gradients and stochastic optimization. The algorithm maintains a current estimate of the global variational parameters, while repeatedly subsampling data. It uses the current global parameters to compute the optimal local parameters for the subsampled data, and adjusts the current global parameters appropriately. 

Implementing a stochastic variational algorithm on the MMSB requires subsampling network data, a difficult problem, especially on sparse networks commonly found in practice. If one subsamples nodes and keeps the links between them, there will hardly be any links in the subgraph, making updates to parameters extremely noisy \citep{ma2017stabilized}. \citet{gopalan2013efficient} instead proposed a sampling strategy that is guaranteed to sample a large number of links often. For instance, their stratified random node sampling selects a node and partitions its node pairs into links and many sets of non-links. The stochastic variational inference selects a node at random and picks a link set or one of its non-link sets half the time. By up-weighting the probability of selecting links, they ensure efficient updates to their global parameters by not allowing their subsampled graphs to be almost filled entirely of non-links at each iteration.

Fitting a MMSB can be improved by developing a sampling procedure that creates subgraphs which more closely resemble the original graph. Although \citet{gopalan2013efficient} create subgraphs with many links, the resulting subgraph does not carry the same properties from the full network. They rely on the subgraph to get noisy, unbiased updates for the global parameters. Although their sampling strategy is superior to naively subsampling nodes, the global parameters recovered from the subgraph will not closely match those in the full graph. Instead, they rely on stochastic optimization to get closer with each iteration towards the truth. Thus, for a fixed node, their procedure necessitates the need to periodically sample subgraphs that contain links involving that fixed node to efficiently update its variational parameter. 


\subsection{Review of ARD}
We show in this paper that employing node-centric aggregated relational data (ARD) gives an efficient solution to the above issue. ARD can be used to create a multigraph using a subsample of nodes that can estimate the global parameters directly, eliminating the need to continuously update the
global parameters
with each sampled subgraph. Aggregated relational data is commonly used in the study of social networks \citep{diprete2011segregation}. Sociologists are interested in the connections between people and gather information through sample surveys \citep{mccarty2001comparing}. Ideally, a respondent would reveal some personal information and enumerate all the persons he or she knows \citep{shelley1995knows}. However, this is not always feasible since people may be reluctant to report membership to a certain group due to social pressure or stigma. 
Additionally, enumerating one's network of friends and acquaintances is not practical especially since personal network sizes run to the hundreds of individuals. To get around this obstacle, survey enumerators ask questions of the form ``How many $X$'s do you know?," where $X$ represents a subpopulation of interest. For example, $X$ can be the subpopulation of people with first name Michael. Rather than having data on the connections individually, we get the total number of links the respondent has with Michaels. This idea can be extended to other networks with well-defined subpopulations. 

\citet{mccormick2015latent} modeled ARD as a partially observed full network. By first positing a model of the complete graph and deriving the model for the ARD, they were able to establish a framework that yielded an explicit relationship between complete graph features and the sampled data. This connection illuminates how inferences made on the smaller graph effects inferences made on the complete graph.

\section{ARDMMSB Model and Algorithm}
\label{sec:model}

In this paper, we address the computational challenge posed by fitting massive real-world networks by proposing ARD to construct bipartie graphs that carry information about full network features. Ideally, we want inferences made
for this bipartite graph to carry over to inferences made on the complete graph. To form ARD, we leverage background information on the dataset to create subpopulations, if it is available. Nodes within a subpopulation should have similar memberships, while the subpopulations themselves ideally should be spread across regions of the membership space. As an example, in a citation network, one can use the journals that the papers were published in as subpopulations. Intuitively, papers in the same journal should have similar community memberships. In this case, for each sampled node, one would summarize the number of citations to each journal to form the ARD bipartite graph. We will show below that this method provides stable estimates of the blockmatrix $B$ and the subsampled nodes' community memberships. However, other methods, such as constructing subpopulations based on papers which share keywords, is also possible.

In what follows, we model the entire full network as a MMSB, create random subgraphs from ARD mini-batches, and model the aggregated links to infer the blockmatrix $B$ and membership vectors for each node in the complete graph. Central to this algorithm is a model for the ARD bipartite graph assuming the underlying full network is generated from an MMSB. 
{\color{red}We first introduce} this model and then describe
our variational inference procedure for ARD, before giving practical remarks
related to implementation.

\begin{figure}[ht!]
	\begin{minipage}{0.35\textwidth}
		\begin{tikzpicture}
		\node[latent]   (pi_i)  {$\pi_i$} ; %
		\node[obs, below = of pi_i, xshift = 2cm] (Y)  {$Y_{ik}$} ; %
		\node[obs, below = of pi_i, xshift = -2cm] (Y2) {$Y_{i\ell}$};
		\node[latent, below = of Y, xshift = -2cm] (B) {$B$};		
				
		\node[latent, above=of Y2]   (pi_gl)  {$\pi_{G_{\ell}}$} ; %
		\node[latent, above = of Y]   (pi_gk)  {$\pi_{G_k}$} ; %
		\node[above=of pi_gl, xshift = 2cm] (a) {$\alpha$};
		
		\edge {a} {pi_i, pi_gl, pi_gk};
		\edge {pi_gl} {Y2};
		\edge {pi_i} {Y, Y2};
		\edge {pi_gk} {Y};
		\edge {B} {Y, Y2};
		
		\plate {} {(B)} {};
		
		\end{tikzpicture}
	\end{minipage}
	\begin{minipage}{0.59\textwidth}
	    \hrule
	    \vspace{2pt}
		\textbf{Generative process of ARDMMSB}
		\vspace{2pt}
		\hrule
			\begin{enumerate}
			    \item For each entry of the blockmatrix $B$, draw probability $B_{ij} \sim \text{Beta}(a_{ij}, b_{ij}).$
				\item For each node $n = 1, \dots, n_b$ in minibatch $b$, draw $\pi_n  \sim \text{Dirichlet} (\alpha).$
				\item For each subpopulation $k = 1, \dots, {\color{red}\kappa}$, draw $\eta_k \sim \text{Dirichlet} (\alpha).$
				\item For each node and subpopulation pair $(i,k)$, draw $y_{ik} \sim \text{Poisson}(N_k \pi_i^T B \eta_k).$
			\end{enumerate}
		\hrule
	\end{minipage}
	\caption{Left: Graphical representation of a two-node segment of the ARD network. The complete model contains $y_{ik}$ for every node, subpopulation pair. Circles denote variables and observed variables are shared. The plates contain variables to be replicated. Right: Data Generating Process for Aggregated Relational Data for MMSB.}
	\label{fig:dgs_ardmmsb}
\end{figure}

\subsection{Proposed model: ARDMMSB}


In a complete graph, we observe an {\color{red}$N \times N$} adjacency matrix
$\boldsymbol{\delta}$
where $\delta_{ij} = 1$ if there is a directed edge from $i$ to $j$ and 0 otherwise. We assume that there are $K$
latent communities present in this network, where the probability of 
an edge between these communities is determined by a $K\times K$ block
matrix $B$, where $B_{lm}$ is the probability of an edge from a node
in community $l$ to a node in community $m$.
In the MMSB, the propensity to form ties is modeled conditionally on $\pi_i$ and $\pi_j$, the memberships of sender $i$ and receiver $j$. The probability is calculated by integrating over the sender and receiver indicators $s_{ij}$ and $r_{ij}$,

\begin{align*}
P(\delta_{ij} = 1 |\pi_i,\pi_j) &= \displaystyle \int P(\delta_{ij} = 1 |s_{ij},r_{ij}) P(s_{ij} | \pi_i )  P(r_{ij}| \pi_j) ds_{ij} dr_{ij} \\
&= \pi_i^T B \pi_j.
\end{align*}

\citet{gopalan2013efficient} infer these global 
parameters $B, \boldsymbol{\pi}$ from subgraphs formed from the 
original graph. While these subgraphs can be constructed to
contain many links, it may still be challenging to 
generate reasonably sized subgraphs which can infer these
global parameters well. For example, when learning 
the off diagonal elements of $B$, the inter group
connection probabilities, there may only be a limited 
number of edges with which to perform inference.
This can lead to noisy estimates of these
parameters, and may make convergence of the inference
scheme challenging.

Instead of observing the connections between each pairs of nodes $(i,j)$, we only observe aggregated counts of links 
{\color{red}
between each node $i$ and each of $\kappa$ subpopulations, such that
$y_{ik} = \sum_{j \in G_k} \delta_{ij}$ where $G_k$ is the $k$th subpopulation, with
$k=1,\ldots, \kappa$.
}
Conditional on the community memberships $\pi_i$ and $\{ \pi_j \}_{j \in G_k}$, $\{ \delta_{ij}\}_{j \in G_k}$ are independent Bernoulli random variables, each with a small probability of success. {\color{red}
It is then} reasonable to assume,
\begin{align*}
    y_{ik} &\sim \text{Poisson}(\lambda_{ik}) \\
   \lambda_{ik} &\approx \sum_{j \in G_k} P(\delta_{ij} = 1 |\pi_i,\pi_j) = \sum_{j \in G_k} \pi_i^T B \pi_j. 
\end{align*}
Since we do not observe $\delta_{ij}$ for $j \in G_k$, we will not be able to estimate the latent parameters $\{ \pi_j \}_{j \in G_k}$ and thus not be able to infer the Poisson rate $\lambda_{ik}$. Instead, we approximate the rate by taking the expectation over the latent positions of nodes in subpopulation $G_k$. 
{\color{red}Defining $N_k$ to be the number of nodes in subpopulation $k$,}
we have the approximation,
\[ 
\frac{1}{N_k}\sum_{j \in G_k} \pi_i^T B \pi_j \approx \expect_{\pi_j \sim P_k} (\pi_i^T B \pi_j)
\]
where we introduced a distribution $P_k$ over nodes $j$ in subpopulation $k$. {\color{red}
This $P_k$ is a distribution over the $K$ communities in the network. This gives}
\begin{equation}
\label{eq:poission_approx}
\lambda_{ik} \approx N_k \expect_{\pi_j \sim P_k} (\pi_i^T B \pi_j).
\end{equation}
The approximation in Equation~\eqref{eq:poission_approx} has two key features. First, the probability of a connection is no longer conditional on the membership of the two nodes but now conditions on the sender's membership and the expected membership of a node in the subpopulation. Second, it introduces a distribution over the set of latent membership vectors, $P_k$, with integration over
{\color{red}the $K-1$ dimensional unit}
simplex.  

Now we must make a choice for the subpopulation distribution. If we take $P_k = \text{Dirichlet}(\alpha_k)$, then
\[ \expect_{\pi_j \sim P_k} (\pi_i^T B \pi_j)  = \pi_i^T B \expect_{\pi_j \sim P_k} (\pi_j) = \pi_i^T B \eta_k
\]
{\color{red} where $\eta_k = (\eta_k^1,\ldots, \eta_k ^K)$ is a vector of length $K$
and
$\eta_k^j =  \left[ \frac{\alpha_k^j}{\sum_{j=1}^{K} \alpha_k^j} \right] $.} 
Combining these results, we have
\[
y_{ik} | \pi_i, \eta_k \sim \text{Poisson} (N_k \pi_i^T B \eta_k ), 
\]
which is the likelihood of the MMSB for ARD. In this paper, we will infer $\eta_k$, the subpopulation mean, rather than $\alpha_k$, which would additionally allow estimation of the subpopulations' concentration. To finalize our model specification we add prior distributions for the community memberships $\pi_i$ and $\eta_k$ as well as the blockmatrix $B$. The data generating processes for MMSB and the Aggregated Relational Data for MMSB is summarized in Figure 2.

\subsection{Posterior Inference}
\label{sec:inference}

To perform inference for large networks, 
we randomly sample {\color{red}minibatches} of nodes and subpopulations and
use the resulting ARD bipartite subgraphs to learn the parameters 
of the underlying MMSM model for the original network.
The variational inference scheme we perform for each of these minibatches is 
described below. 

As the true posterior of our model is not available in closed form, we develop an efficient variational algorithm for posterior approximation.  
We fit this procedure 
to each of the minibatches independently. This allows us 
to parallelize this algorithm across minibatches. 
{\color{red}These are minibatches of ARD data, namely samples of nodes and the
corresponding subpopulation counts. This retains more 
expressive 
information than minibatches of the underlying adjacency matrix, $\boldsymbol{\delta}$.}

Let $\Theta$ denote the set of unknown variables in the ARDMMSB,
{\color{red} namely the community memberships, $\pi$, the subpopulation means, $\eta$,
and the underlying blockmatrix, $B$}. In variational inference, the true posterior is approximated by tractable distributions which are optimized to be close to the true posterior in terms of Kullback-Leibler divergence. Here we consider a fully-factorized family (commonly called mean field variational inference),

\[
q(\Theta) = \prod_{i,k} q_K(\pi_i | \gamma_i) q_K(\eta_k | \phi_k) \prod_{m,n} \delta_{B_{mn}} 
\]

{\color{red} where we infer a separate variational family for each of the latent
variables in our model. For both $\pi_i$ and $\eta_k$ a natural choice is to consider
a Dirichlet variational family, with parameters $\gamma_i$ and $\phi_k$ respectively.
We use $q_K$ to denote these Dirichlet distributions of order $K$,} 
while $ \delta_{B_{mn}}$ is the point mass at $B_{mn}$. 
{\color{red}In variational inference we aim to infer the variational parameters such 
that the variational approximation best matches the true posterior distribution.
We aim to optimise over the variational parameters, $\{\gamma, \phi, B\}$ 
to minimize the Kullback-Liebler divergence between the variational approximation, $q(\Theta)$,
and the true posterior.}

From Jensen's inequality, minimizing the Kullback-Leibler divergence between $q(\Theta)$ and the true posterior is equivalent to maximizing a lower bound $\mathcal{L}$ on the log marginal likelihood \citep{blei2017variational}. 
{\color{red}The lower bound is given by}
\begin{align}
\mathcal{L} &= \expect_q \log p(Y, \Theta) - \expect_q \log q(\Theta) \label{eq:vi_lower_bound}\\
& = \sum_{i,k} \expect_q \log p(y_{ik} | \pi_i, \eta_k, B) + \sum_i \expect_q \log p(\pi_i | \alpha) \nonumber \\
&\quad + \sum_k \expect_q \log p(\eta_k | \alpha) + \sum_{m,n} \expect_q \log p(B_{mn} | a_{mn}, b_{mn}) 
+ \mathcal{H}(q). \nonumber
\end{align}

{\color{red}
Here note that $\expect_q \log p(Y, \Theta)$
in ~\eqref{eq:vi_lower_bound} consists of both the priors for $\pi,\eta$ and
$B$, along with the likelihood of the MMSB for ARD. 
We consider a common Dirichlet prior with $K$ dimensional 
parameter $\alpha$ for $\pi$ and $\eta$. For entry $B_{mn}$
of $B$ we utilise a Beta$(a_{mn},b_{mn})$ prior. In practice, we 
choose $a_{mn}=b_{mn}=1$, giving a uniform prior on the entries
of $B$, and a common concentration vector for the Dirichlet prior
$\alpha = (\alpha,\ldots, \alpha)$.
}

However, $\expect_q \log p(y_{ik} | \pi_i, \eta_k, B)$ cannot be evaluated in closed form. To circumvent this issue, we lower bound this term further by introducing auxiliary parameters.
\begin{align}
\expect_q & \log p(y_{ik} | \pi_i, \eta_k, B) \nonumber \\
&= y_{ik} \expect_q \log(\pi_i^T B \eta_k) + y_{ik} \log N_k - N_k \expect_q \pi_i^T B \eta_k \nonumber \\
&\geq y_{ik} \sum_{m,n} \p \expect_q \log(\pi_i^m B \eta_k^n) - \sum_{m,n} \p \log \p \nonumber \\ 
&\quad + y_{ik} \log N_k - N_k \expect_q \pi_i^T B \eta_k. \label{eq:lower_bound}
\end{align}

where $\{ \p | m,n = 1, \dots, {\color{red}K} \}$ is an auxiliary probability vector for every $(i,k)$ pair. The lower bound 
on the log marginal likelihood 
obtained by using Equation~\eqref{eq:lower_bound} is denoted by $\mathcal{L}^*$. 
{\color{red}The full expression for $\mathcal{L}^*$ is included in the appendix.}

We optimize $\mathcal{L}^*$ via coordinate ascent, {\color{red}
optimizing both the variational parameters $\{\gamma, \phi, B\}$ and the corresponding
auxiliary parameters $p_{ik}^{(mn)}$}.
For the auxiliary parameters $\p$, we update $\mathcal{L}^*$ by tightening inequality \eqref{eq:lower_bound}. For $\{\gamma, \phi \}$, the likelihood is nonconjugate with respect to the prior. We appeal to nonconjugate variational message passing for updates of these parameters \citep{knowles2011non}. This is a fixed point iteration method for optimizing the natural parameters of variational posteriors in exponential families. The advantages of this approach is that it yields closed form updates and extends to stochastic variational inference naturally. However, $\mathcal{L}^*$ is not guaranteed to increase at each step and updates for $\{ \gamma, \phi \}$ may be negative at times. To resolve these issues, we use the fact that nonconjugate variational message passing is a natural gradient ascent method with step size 1 and smaller step sizes may also be taken. {\color{red} An outline of the overall procedure is
shown in Algorithm~\ref{Algorithm 2}.}
{\color{red}When updating $\{ \gamma, \phi \}$ using nonconjugate 
variational message passing, }
we start with step size 1 and reduce the step size where necessary to ensure updates
are positive. If $\mathcal{L}^*$ increases, these updates are accepted. Otherwise, we revert to 
the former values. 
{\color{red}As a point mass $\delta_{B_{mn}}$ 
was chosen for the variational family for $B$, the entries
are updated by a standard gradient step. We show the expression for the updates
of
$\{ \gamma, \phi \}$ in Algorithm~\ref{Algorithm 2}. The derivation of these
updates and their exact form is
included in the Supplementary Material.}


\begin{algorithm}[ht]
\small
\begin{flushleft}
Initialize
{\color{red} the variational parameters}
$\gamma$, $\phi$, $B$. Cycle through the following steps until all nodes are sampled.
\end{flushleft}

\begin{enumerate}
	
\item Randomly sample $n$ nodes without replacement,
{\color{red} obtaining sample $\mathcal{S}$.}

\item Randomly sample $k$ {\color{red}of the $\kappa$} subpopulations without replacement.

\item Initialize local {\color{red}
auxiliary probability}
variables $p$
{\color{red} and step size $s_t=1$}. 
Cycle through
steps 4 to 8 until convergence.

\item Update $B$ {\color{red} using a gradient step, $\hat{B}_{mn} = \frac{\sum_{i,k} y_{ik}p_{ik}^{(mn)}}{\sum_{i,k} N_k \frac{\gamma_i^m}{\sum_d \gamma_i^d} \frac{\phi_k^n}{\sum_d \phi_k^d}}$ for a uniform prior for $B_{mn}$.}

\item Update $\gamma_i \leftarrow (1 - s_t)\gamma_i + s_t \hat{\gamma}_i$ for $i \in \mathcal{S}$ where
$\hat{\gamma}_i = \textit{I}^{-1}_{\gamma_i} \nabla_{\gamma_i} \expect_q [\log p(y, \Theta)].$
If any element of $\gamma_i \leq 0$, reduce $s_t$ (say by half each time) until $\gamma_i > 0$. Accept update only if $\mathcal{L}^*$ increases.

\item Update $p$ {\color{red} by tightening the lower bound $\mathcal{L}^*$, $p_{ik}^{(mn)} \propto \exp \left [ \expect_q \log(\pi_i^m B \eta_k^n) \right ]$}

\item Update {\color{red}
$\phi_l \leftarrow (1 - s_t)\phi_k + s_t \hat{\phi}_l$ for $l \in 1, \dots, K$ where
$ \hat{\phi}_l = \textit{I}^{-1}_{\phi_l} \nabla_{\phi_l} \expect_q [\log p(y, \Theta)]. $
If any element of $\phi_l \leq 0$, reduce $s_t$ (say by half each time) 
until $\phi_l > 0$. Accept update only if $\mathcal{L}^*$ increases.}

\item Update $p$ {\color{red} given these
variational estimates, $p_{ik}^{(mn)} \propto \exp \left [ \expect_q \log(\pi_i^m B \eta_k^n) \right ]$}


\item Average the subpopulation parameters $\phi_l$ and blockmatrix $B$ across minibatches.

\end{enumerate}
\caption{Variational inference procedure for ARD mini-batches}
\label{Algorithm 2}
\end{algorithm}

\begin{figure}[ht!]
    \begin{center}
	\includegraphics[width = 0.8\textwidth]{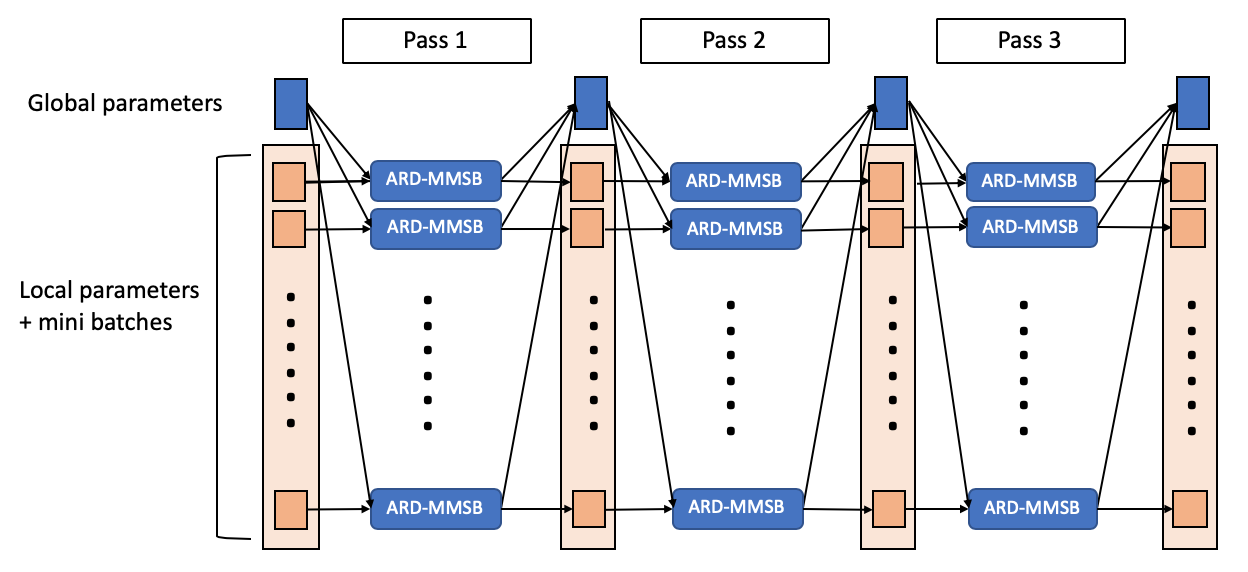}
    \end{center}
	\caption{Illustration of inference process for multiple passes. Each tall orange rectangle represents all of the variational parameters for the nodes. The blue blocks represent the subpopulation and blockmatrix parameters while the orange blocks represent the parameters for each node. In each pass, the orange blocks are broken up into minibatches. Each of the minibatches are passed along with the current blue parameters and fit through the algorithm. After the pass, the orange blocks are stored and the blue parameters are averaged over before 
    being stored.}
	\label{fig:algo}
\end{figure}


\paragraph{Multiple Passes.}
When fitting to a large network, each minibatch will contain a small fraction of nodes. 
After initialization, the nodes in each minibatch will be run through the algorithm with 
weakly informative subpopulation blockmatrix parameters. The fit of each node ignores 
link information from all other minibatches. However, after being fit with 
Algorithm~\ref{Algorithm 2}, the subpopulation and blockmatrix parameters contain richer 
information since they are averaged over all minibatches. Running 
Algorithm~\ref{Algorithm 2} a second time with the fits as initial values will allow the 
node parameters to be fit using information across the network. This process is 
summarized in {\color{red}Algorithm S1 of the supplementary material}.
We found that two passes is usually sufficient for the stability of parameters. This entire process is illustrated in Figure~\ref{fig:algo}.  

		
		
		
		
				

\paragraph{Per-iteration Complexity.}
At first glance, it seems that we must store a matrix of ancillary parameters 
$[p_{ik}^{(mn)}]_{m,n}$ for each node-subpopulation pair. This would mean 
having a memory requirement of $O(D^2NK)$. This can be problematic when trying 
to fit the model with a large number of communities. However, all parameter 
updates depend on $p_{ik}^{(mn)}$ only through the term $y_{ik}p_{ik}^{(mn)}$. {\color{red}Thus,}
we can take advantage of the sparseness found in real-world networks by only storing ancillary parameters corresponding to nonzero counts $y_{ik}$.

\paragraph{Picking subpopulations.}
Ideally, subpopulations should be chosen so that the members 
of a subpopulation
have similar community membership. In practice, one would need to use background information to pick subpopulations \citep{RN2, RN4, RN5},
{\color{red} which may not be readily available.}
For instance, in a citation network, journals may be a good choice of subpopulations, as papers within a journal are often on the same topic. 
This will result in non-overlapping subpopulations.
Alternatively, we could use specific keywords as a choice of subpopulations,
which may result in overlapping subpopulations. For the simulation 
studies demonstrating the performance of our proposed method in 
Section~\ref{sec:simulation} there is no background information 
and so the subpopulations were randomly chosen and still resulted 
in strong community recovery.
Similarly, we note that while there 
is a computational cost to initially forming these subpopulations counts 
of $\mathcal{O}(n^2\kappa)$, this calculation is only required initially
and does not need to be repeated.

\paragraph{Initialization}
From our synthetic data set, we found initialization of $\{ \gamma, \phi \}$ is 
important for good recovery of the model parameters. We initialized by first forming 
a new matrix $\tilde{Y} = [y_{ik}/N_k]_{ik}.$ That is $\tilde{Y}$ is a normalized 
version of the ARD. Communities for the sampled nodes were initialized using soft 
clustering on the top {\color{red}$K$} left singular values {\color{red} of $\tilde{Y}$},
while the subpopulation communities were clustered using the top {\color{red}$K$} right singular values. 
{\color{red}We use this initialization scheme for all 
comparison methods considered
in the simulation studies and real data example below.}

\section{Simulation Studies}
\label{sec:simulation}

In this section we perform simulation studies showing that bipartite 
graphs formed from ARD preserve enough information from the full network so that inference made 
using ARD minibatches carries over to inference based on the complete graph. Moreover, we compare our method to 
\citet{gopalan2013efficient} and observe 
{\color{red} that we have excellent parameter 
recovery and model fit using ARD subgraphs, without observing all nodes in the network.
The procedure of \citet{gopalan2013efficient}
is unable to correctly recover the true communities or model parameters using only a subgraph 
of the original network, only achieving comparable performance when the complete network
is used.}


\subsection{Subsampling}
\label{subsec:subsampling}

{\color{red}
In practice, a network is often presented to the researcher that is too large to conduct an 
analysis with the complexity required to gain an insight into the underlying network.}
Fitting Bayesian hierarchical models to such data simply take too much time. Because of the intractability of such models, the researcher oftentimes runs her analysis on a subgraph of the network. Choices of subgraphs include the largest connected component or a subset of nodes with similar covariates. However, inferences on the subgraph may be misleading when extended to the entire network. Ideally, the subgraph will in some way be representative of the original graph. Among the many sampling strategies of networks, simple random node selection has been shown to create subgraphs that maintain many features of the graph \citep{leskovec2006sampling}.

If the complete graph is a MMSB, subsampling nodes and doing inference on the resulting 
subgraph may lead to misleading inferences. 
Due to its data generating process, sampling nodes at 
random and keeping the edges between them will result in a MMSB with the same blockmatrix. 
{\color{red}However,} such a procedure may lead to a large loss in efficiency and thus 
unstable parameter estimates. This is particularly an issue in sparse networks since such a 
subsampling procedure will leave out most of the links in the network,
making it hard to learn the many parameters of the model. Using ARD leads to vastly improved 
estimates of both the blockmatrix and membership profiles.

\begin{figure*}[ht]
	\begin{minipage}{0.49\textwidth}
	    \centering
		\includegraphics[width=\textwidth]{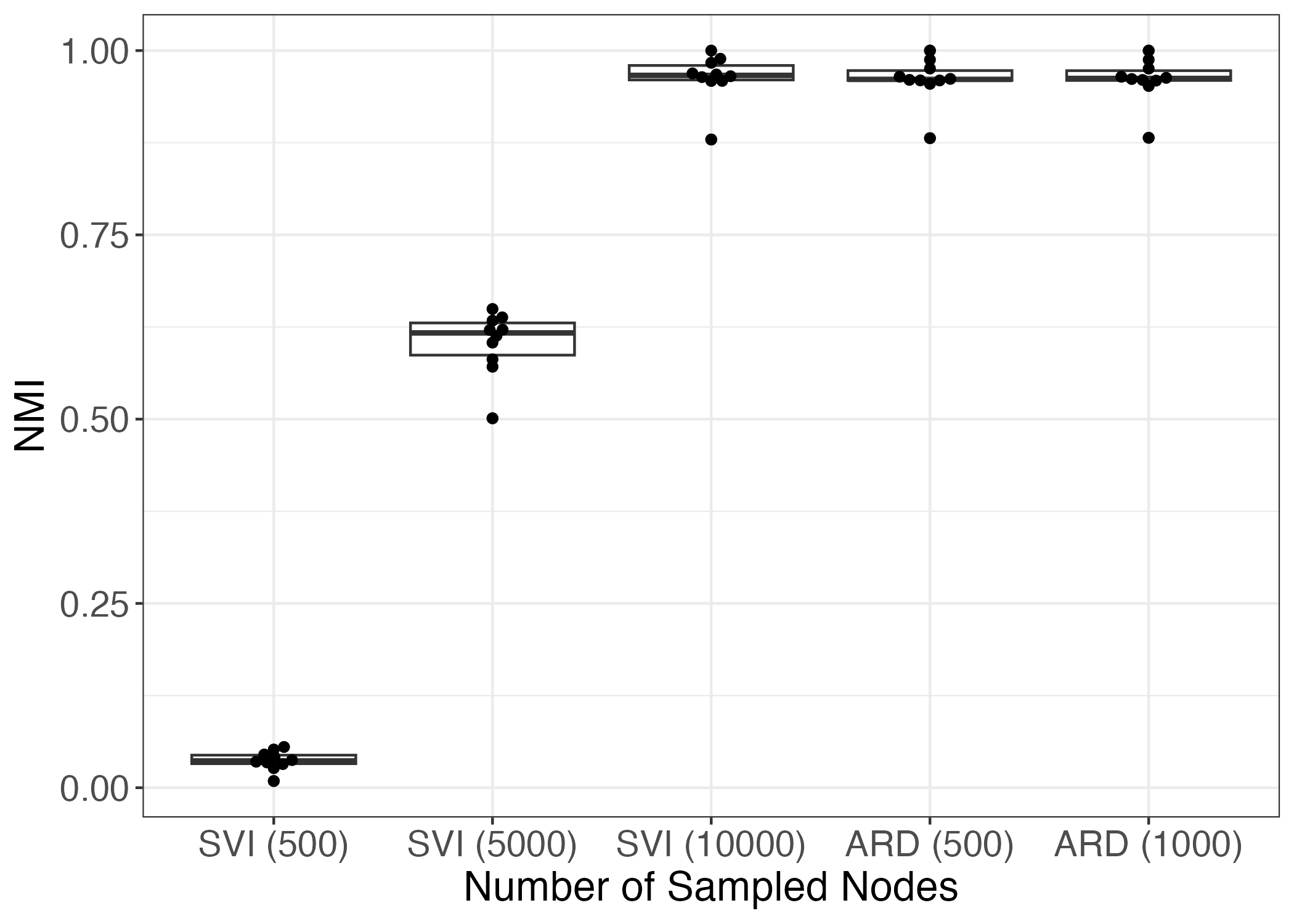}
	\end{minipage}
	\begin{minipage}{0.49\textwidth}
	    \centering
		\includegraphics[width=\textwidth]{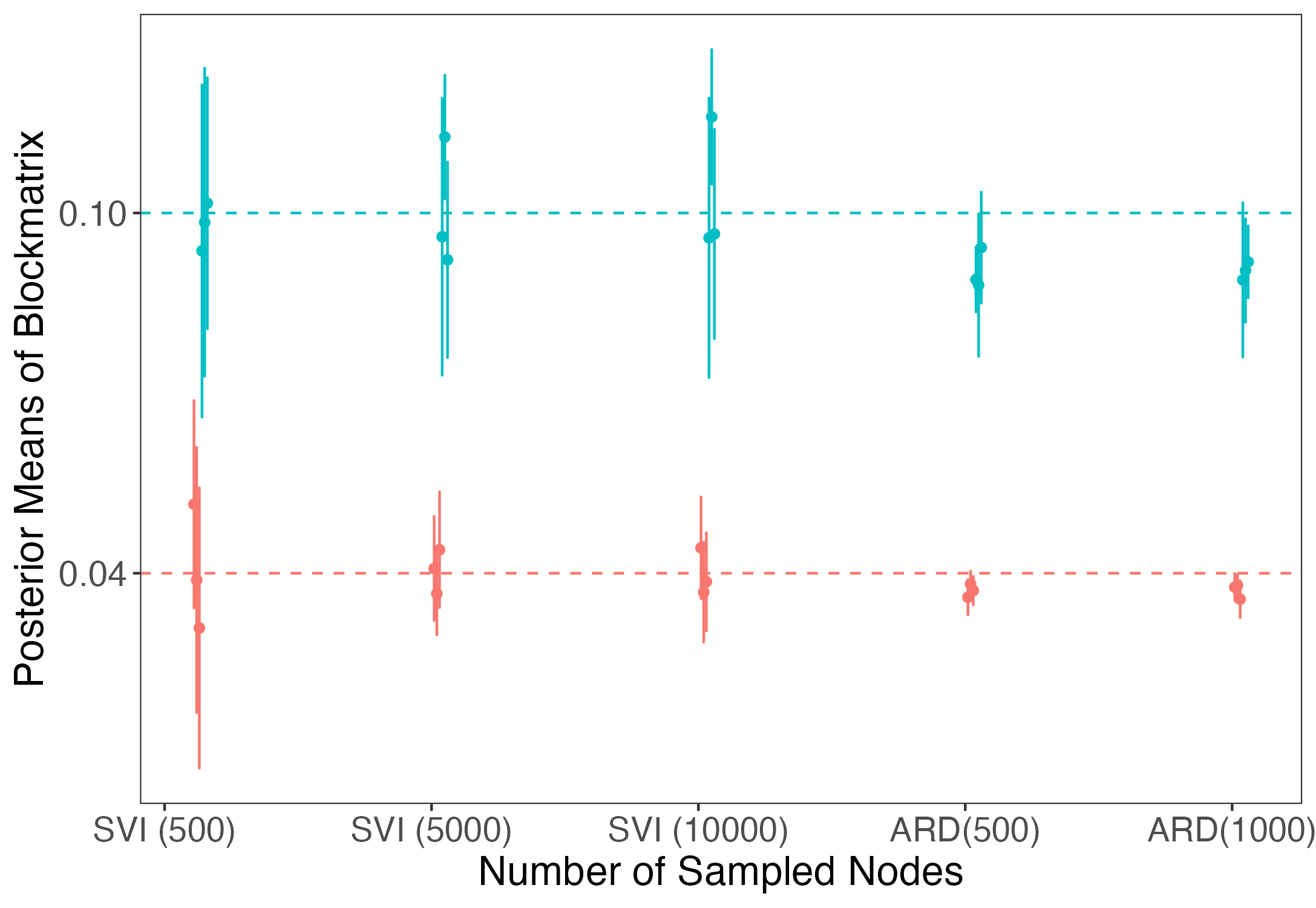}
	\end{minipage}

	\caption{Left: Boxplots of normalized mutual information (NMI) among the subgraphs considered in simulations, using SVI for subgraphs of size $n$
    of the underlying adjacency matrix and ARD subgraphs of size $n$. Right: Posterior means and standard errors of estimation of diagonals of blockmatrix for SVI on subgraphs
    of the adjacency matrix and ARD data of size $n$.
    The true values of the block matrix are given as dashed horizontal lines. }
	\label{fig:sim1}
\end{figure*}

We first wish to investigate the choice of subsample size on 
inference for such network data. {\color{red}Can valid inferences about the 
overall network structure be obtained using ARD subgraphs, rather than 
the whole network? Is this possible with existing methods for MMSB data, 
taking subgraphs of the underlying adjacency matrix instead?}
Figure~\ref{fig:sim1} shows the results from a simulation experiment illustrating the instability 
of estimates from a subgraph formed by sampling nodes. 
We evaluate performance by community assignment and blockmatrix recovery. 
In this experiment, we 
simulated 
{\color{red}10 networks, each a MMSB with $N=10000$ nodes and $K=6$ communities.}
Our model assumes that we have 
information other than the link structure that can define subpopulations or groups of similar 
nodes. Thus, for our simulations, we must introduce a method of generating these subpopulations. 
We generated {\color{red} $\kappa=50$}
subpopulation centers and then generated the members 
{\color{red} or each subpopulation}
from a Dirichlet at 
the subpopulation center. The blockmatrix {\color{red}$B$} 
is diagonally dominated to ensure exact recovery is 
possible with enough nodes \citep{zhao2012consistency}, {\color{red} with non-diagonal entries 
being 0.}
Three communities have within-linking probabilities of 0.1 while the other three have within-
linking probabilities of 0.04. 
{\color{red}Given these
simulated networks, we form subgraphs of the original network first, 
sampling $n = 500$ and $n=5000$ nodes uniformly at random and keeping the edges between 
them. }
{\color{red}Similarly, ARD subgraphs were}
formed with $n=500$ and $n=1000$ sampled nodes. 
{\color{red}
Throughout, the use of ARD corresponds to fitting our ARDMMSB model to subgraphs of ARD
data using the inference scheme described in Section~\ref{sec:inference}.
We wish to examine the performance of these different network subgraphs (both original network
subgraphs and ARD subgraphs)
in terms of community and parameter recovery. To investigate how these
samples compare to using the complete network,}
we also include the results 
from fitting the entire network using the stochastic variational inference scheme 
of \citet{gopalan2013efficient}.
We adapt the stochastic variational inference algorithm presented in \cite{gopalan2013efficient} to fit the subgraphs formed from random node sampling and using the entire network,
{\color{red}
(with SVI indicating the use of this inference scheme throughout this paper)}.
Initialization of the nodes' variational parameters was performed using ten random restarts of spectral clustering. We report the best fit for each of the random restarts. 

The left panel of Figure~\ref{fig:sim1} is a boxplot demonstrating performance of community recovery for the different subgraphs. We chose normalized mutual information between the fitted network's membership profiles and the true membership profiles as the measure of performance \citep{danon2005comparing}. For each node, we take its community assignment to be the maximum community in its membership profile. {\color{red}We show the result for $n=500,5000$
for subgraphs of the original network, using SVI and ARD subgraphs of size $n=500,1000$.}

The bipartite graphs formed using ARD have excellent recovery, 
comparable to using the entire network {\color{red}(SVI($10000$))}.
Using SVI with subgraphs
rather than the full graph
struggles to recover the communities well. The right panel of Figure~\ref{fig:sim1} plots 
the average posterior mean of the blockmatrix $B$'s diagonal elements.
{\color{red}We show these for ARD (ARD (500), ARD (1000)) and node subgraphs
(SVI (500), SVI (5000)), along with using the entire network (SVI (10000))}.
As expected, we see that for the subgraphs formed from random sampling, the estimates have little bias and the standard error decrease as we increase the size of the subgraph, although it is still large with even large subgraphs. ARD incurs similar bias but has drastically smaller standard errors. {\color{red}Thus,} ARD contains much more information about the original graph as it performs just as well as fitting on the entire network. The small bias of the ARD
estimates here is related to the validity of the Poisson approximation, which
may not be reasonable for the large subpopulations.

\begin{figure*}[ht]
    \centering
	    \centering
		\includegraphics[scale = 0.4]{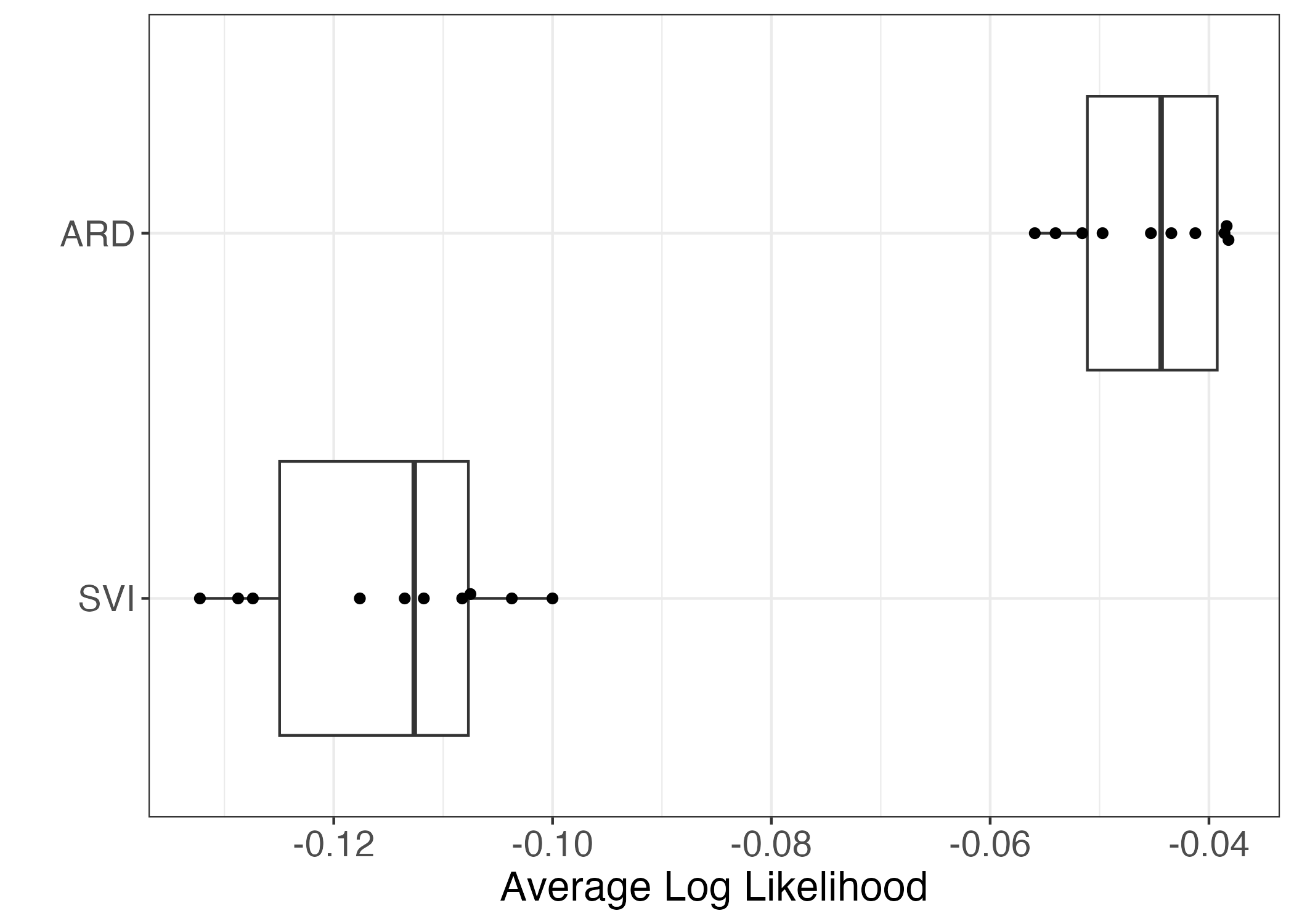}
	\caption{Comparison between ARDMMSB (ARD) and Gopolan \textit{et. al.}'s Stochastic Variational Inference (SVI), 
    showing the average predictive log likelihood with subgraphs of size $n=500$.
    }
	\label{fig:sim2}
\end{figure*}

\subsection{Comparison to Related Method}
\label{subsec:comparison}

Expanding on the results showing the strong performance of our 
ARD procedure with small subsamples, 
we further compare our model and inference algorithm to the SVI algorithm in 
\citet{gopalan2013efficient}. Their method implements a stochastic variational inference 
algorithm for the MMSB. 
{\color{red}
We utilise the same simulation setting}
discussed 
in section~\ref{subsec:subsampling},
{\color{red} generating networks of 10000 nodes.}
For a fixed subgraph size (500 nodes, 5\% of the total network),
we wish to compare the performance of these
two procedures.
We do this in terms of model convergence and community membership recovery. The results are shown in Figure~\ref{fig:sim2} and Figure~\ref{fig:sim1}.

{\color{red}While the previous results showed that SVI required the complete
network to achieve similar community and parameter recovery to ARD subgraphs,
we also wish to quantify
how well our algorithm's solution fits the data compared to that of stochastic 
variational inference. }
Ideally, this would mean comparing the ELBOs, the target criterion of 
variational inference algorithm. Comparisons of the respective ELBOs is not possible since 
\citet{gopalan2013efficient} implement a stochastic variational inference and therefore do not 
store any local parameters. Without the local parameters, the ELBO cannot be computed. Moreover, 
even if we could calculate the respective ELBOs, they would not be comparable since the models 
are different.

Since the ELBO cannot be computed when implementing stochastic variational inference,
\citet{gopalan2013efficient} evaluate model fitness through the predictive distribution \citep{geisser1975predictive}. 
Intuitively, a better model will have a higher predictive likelihood on a held-out set. The held-out predictive likelihood is 
thus used as a proxy for the ELBO. This is computed as follows:
\begin{align*}
    p(&y^{\text{new}}_{ab} = 1| y^{\text{obs}}) \\
    &= \int \theta_a ^T B \theta_b p(\theta_a, \theta_b, B | y^{\text{obs}}) d\theta_a d\theta_b dB \\
    &= \int \theta_a ^T B \theta_b p(\theta_a| \theta_b, B, y^{\text{obs}}) p(\theta_b, B | y^{\text{obs}}) p(B | y^{\text{obs}}) d\theta_a d\theta_b dB \\
    &\approx \int \theta_a ^T B \theta_b q(\theta_a) q(\theta_b) q(B) d\theta_a d\theta_b dB = E_q \theta_a^T E_q B E_q \theta_b. 
\end{align*}

We could also compute a predictive likelihood for our model, but they still would not be 
comparable. However, our algorithm yields a variational posterior for the node memberships and 
the blockmatrix. We can use these parameter estimates and plug them into the posterior likelihood 
defined above. Ideally, both of our models should be fit using the same held-out set of node 
pairs; but it is not obvious how the ARDMMSB model can handle such held-out node-pairs. Instead, we 
propose computing the predictive likelihood over the observed data. A good model fit should have 
a high predictive likelihood on the training set. 
{\color{red} We show the predictive likelihood over the observed data, using ARD subgraphs of size
500 and SVI with subgraphs of size 500 in Figure~\ref{fig:sim2}.}
Figure~\ref{fig:sim2} shows that ARD achieves a higher model fit {\color{red}than SVI using a subgraph
of the original network with the same number of nodes}.

{\color{red}
Due to space constraints, we defer further detailed comparisons
to the supplementary material. There, we investigate our proposed method
as we vary the number of communities,  subpopulations, and sparsity of the 
network. We also examine the computational performance relative to 
SVI.
}
\section{Applications}
\label{sec:empirical}


\begin{figure}[ht]
	\centering
   \includegraphics[width=0.45\textwidth]{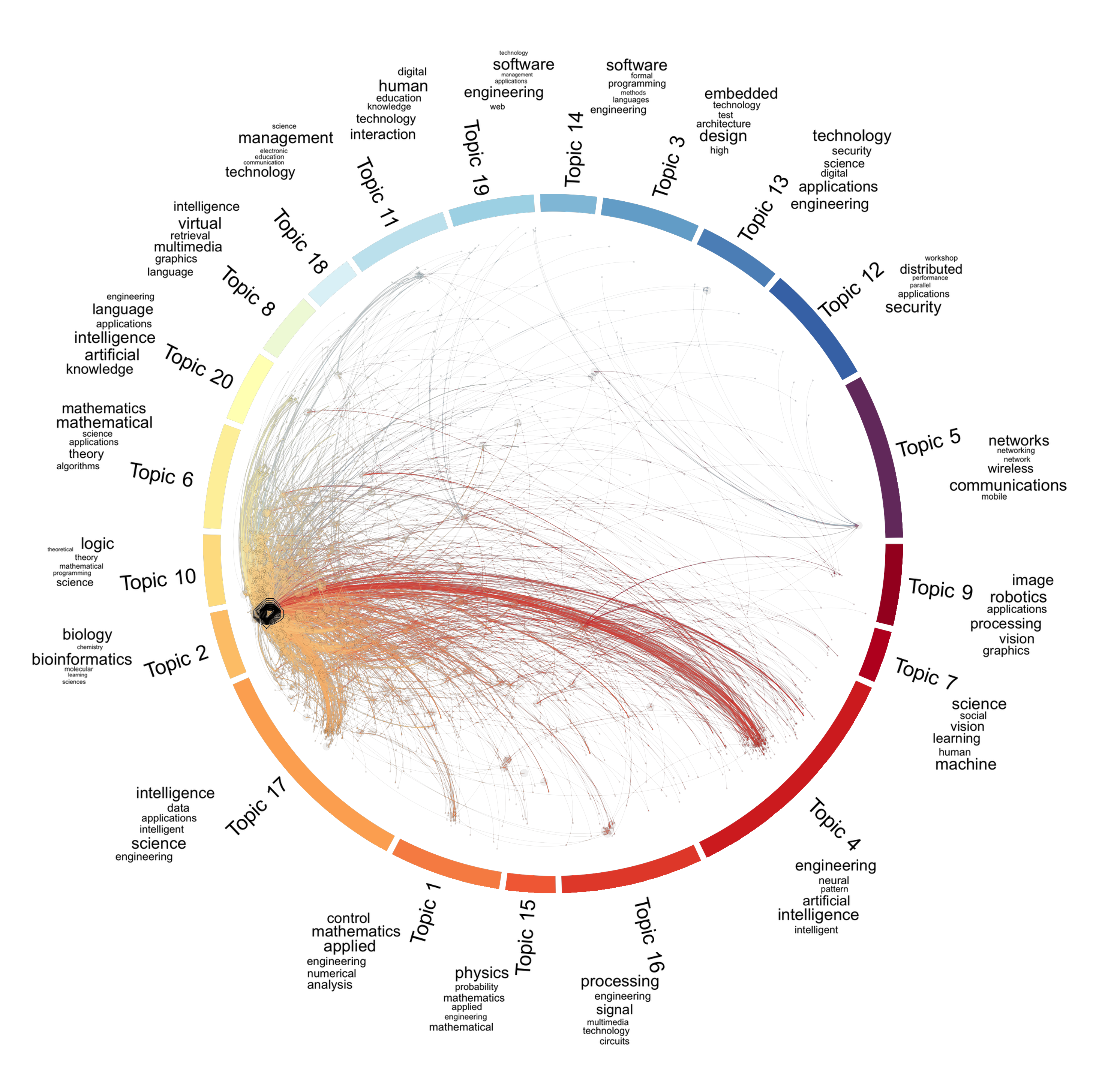}
       \includegraphics[width=0.45\textwidth]{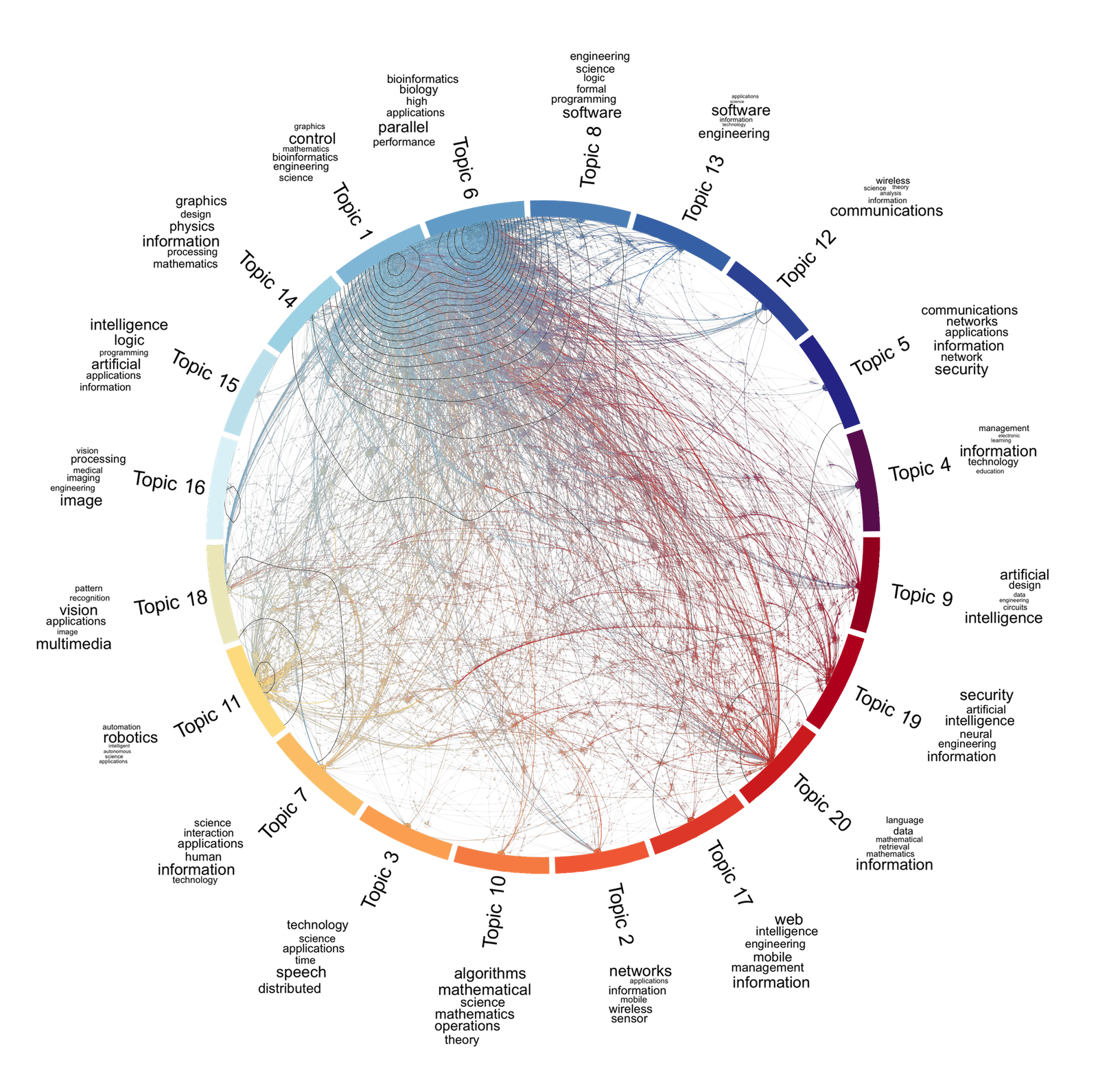}
	\caption{Left: Topic terrain plot of Bioinformatics using results from ARDMMSB. Right: Topic terrain plot of Bioinformatics using results from \citet{gopalan2013efficient}'s stochastic variational inference. A word cloud of the top seven words is displayed with each topic. The font size is proportional to the term frequency.}
	\label{fig:emp1}
\end{figure}

We fit a MMSB using our algorithm to a real-world network to demonstrate how it can help study massive networks. We analyzed a citation network with over three million nodes and 25 million edges extracted from DBLP, ACM, MAG, and other sources \citep{tang2008arnetminer}. We removed journals and papers that are isolated or do not have any outlinks. This reduced network has 2,139,891 papers and 4,349 journals.

For this network, we set the number of communities to twenty and used 1,000 mini-batches. The number of mini-batches used was chosen simply for convenience as the cluster used for fitting took a maximum of 1,000 jobs at once. In each minibatch, we used all the journals and a random subset of roughly 2,000 papers. We measure convergence according to our lower bound approximation of the ELBO. We stopped each mini-batch computation when the change in the ELBO was less than $10^{-2}$.

To initialize the community membership profiles, we formed the journal to journal adjacency matrix and performed regularized spectral clustering. The result gave us hard community assignments for each journal. Each paper's membership was then initialized according to its journal's hard assignment. With this initialization our ARDMMSB model was used to obtain mixed memberships for each of the papers and journals.

Figure~\ref{fig:emp1} shows our community detection results from the fitted model. Each panel shows what we call a topic terrain plot. This plot visualizes the breath and coverage of journals in the citation network. Each plot contains a circle plot of a particular journal. The circle is outlined with twenty colored blocks that represent the twenty communities that papers belong to. The size of a block corresponds to the size of the community or the number of papers in that community determined by the dominant community in its membership profile vector. The placement of the bars around the circle is the result of hierarchical clustering of the twenty topics using inverse value of the probability of a link as the distance metric. Each topic also has a word cloud of the seven most frequent words found among the journal titles after removing stop words.

Inside the circle, we plot the papers' membership profile vectors within that journal as well as the coordinates of papers cited by papers within the journal. We randomly sampled 1000 papers from the journal to construct the plot. For each paper, we sample approximately fifteen of its cited papers on average. The bigger nodes within the wheel represent the papers from the journal and the smaller nodes represent the cited papers. Each link represents a citation activity. The coordinate of each node is the average of the coordinates of the centers of the twenty topic bars, weighted by the membership value of the node. The color of each node is also the average of the RGB value of twenty topic bars, weighted the same way. The color of each edge depends on the target node. The contour plot represents the density estimate of the papers from the journal.

From the word clouds, we can infer what communities correspond to which fields and topics. Those in yellow are related to mathematics. Those in orange correspond to biology and other biological applications, while those in light blue have topics relating to software and software design. Machine learning and artificial intelligence dominate the red region.

The top of Figure~\ref{fig:emp1} shows the topic terrain plot for Bioinformatics using results from our algorithm. Intuitively, we would expect papers in this journal to develop technologies for biological applications. {\color{red}Thus,} they should cite papers in fields such as machine learning and computer vision as well as biology papers which is the citation behavior we see in the figure. We immediately notice that this journal has a strong footprint and unique identity. Most of the papers within Bioinformatics have a high membership in topic 2, which corresponds to biology and biological related fields. Many of papers in published in Bioinformatics cite papers with high membership in topic 4 which correspond to artificial intelligence and machine learning. There are also a significant number of papers being cited in fields such as mathematics and logic.    

The bottom of Figure~\ref{fig:emp1} shows the topic terrain plot for Bioinformatics using results from \citet{gopalan2013efficient}'s stochastic variational algorithm. First notice that the word clouds are much more homogeneous, indicating their algorithm has difficulty detecting meaningful communities. Also, most of the papers belong exactly to one community; that is, papers do not exhibit mixed membership. This could be due to the fact that the topics themselves are not very distinct from one another or the algorithm has difficulty capturing interdisciplinary papers.  


\begin{figure}[ht]
	\centering
   \includegraphics[width=0.45\textwidth]{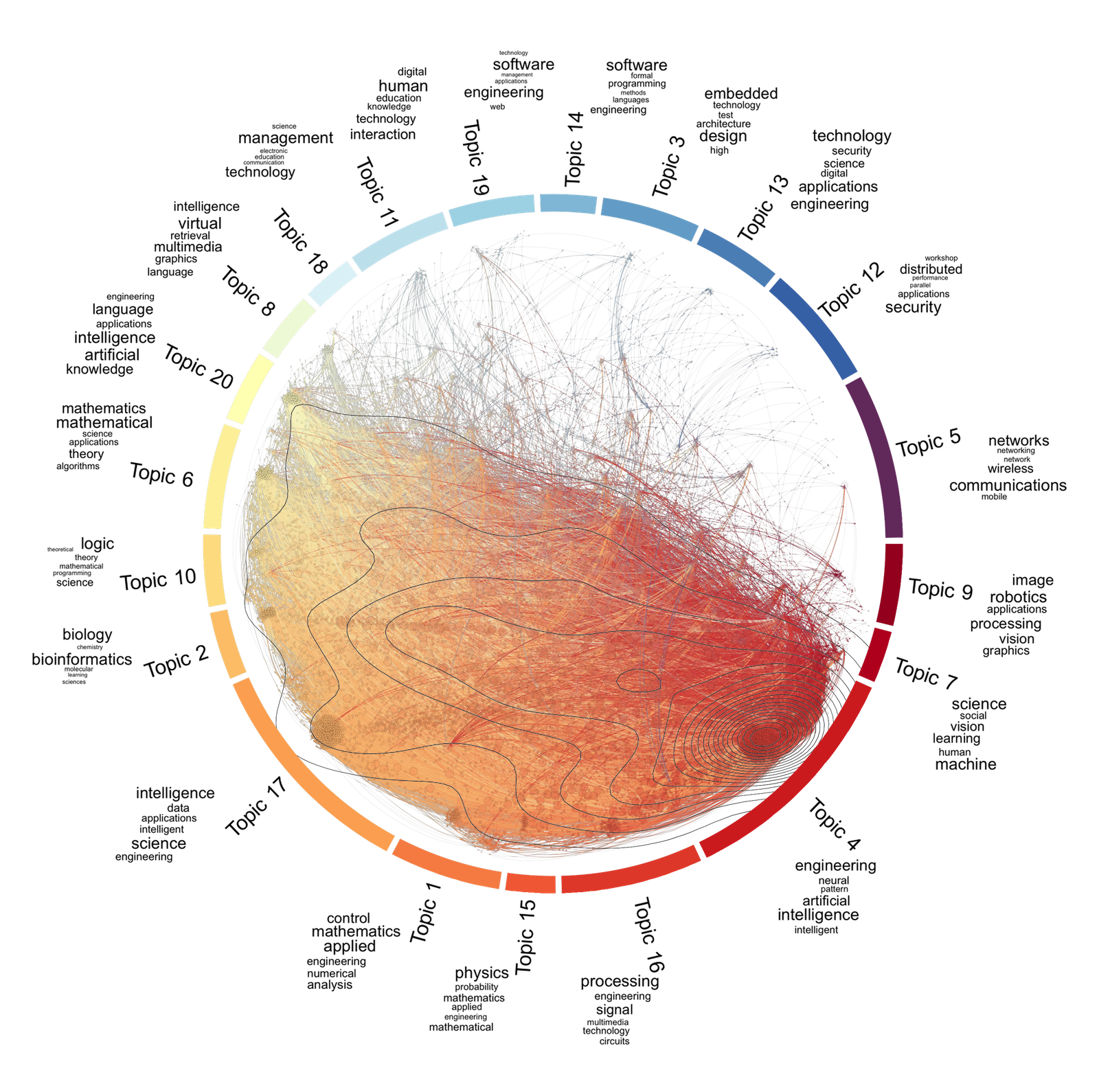}
       \includegraphics[width=0.45\textwidth]{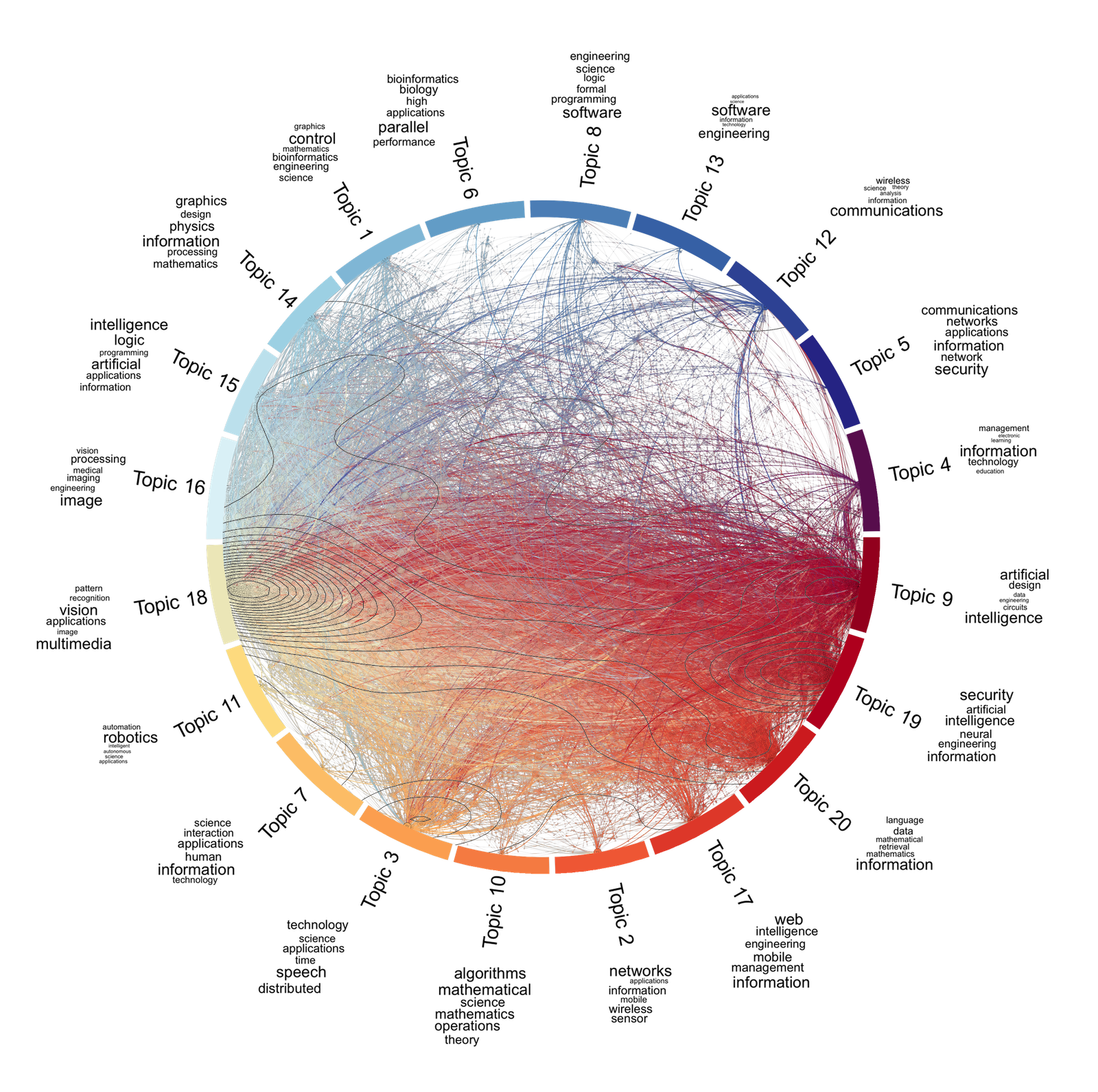}
	\caption{Left: Topic terrain plot of JMLR using results from ARDMMSB. Right: Topic terrain plot of JMLR using results from \citet{gopalan2013efficient}'s stochastic variational inference. A word cloud of the top seven words is displayed with each topic. The font size is proportional to the term frequency.}
	\label{fig:emp3}
\end{figure}

Figure~\ref{fig:emp3} shows the topic terrain plots of the Journal of Machine Learning and Research (JMLR) from the two algorithms. Although JMLR is a journal in machine learning, we expect papers to be broadly spread out among many communities since machine learning can be organized into smaller communities within its broad research landscape. It covers natural language processing, mathematics, logic, machine learning with biological applications, control theory, physics, vision and robotics. In the top plot of Figure~\ref{fig:emp3}, the half circle represents the foundations and applications of machine learning. Topic 4 has the highest concentration which contains journals based in artificial intelligence (AI) and neuroscience inspired AI, such as neural networks. Moreover, there is a large concentration in topic 17 which corresponds to AI without neuroscience. This agrees with our view of ML as a field. It confirms our method is recovering meaningful structure in the citation network.

The bottom plot in Figure~\ref{fig:emp3} continues the trend of having all the papers stuck on the edges, meaning that most papers belong to solely one topic. {\color{red}Thus,} the algorithm cannot capture the interdisciplinary nature of many of the computer science papers that have applications in other areas. Also, the topics themselves do not have strong separation between fields. For instance, computer systems and ML Topics are mixed together and are dominated more by  application areas rather than research areas. 

By comparing the two plots, we illustrate how integrating nodal information can greatly improve results on real-world networks. For JMLR, our method divides the topic terrain plot into two half circles, one half being ML and the other about hardware, system design, security, etc. This is due to incorporating journal information which is essentially human-curated structure. Leveraging extra information allows ARDMMSB to uncover more structure in the citation network.

\begin{figure}[!htbp]

\begin{minipage}{0.45\textwidth}
\includegraphics[width=\linewidth]{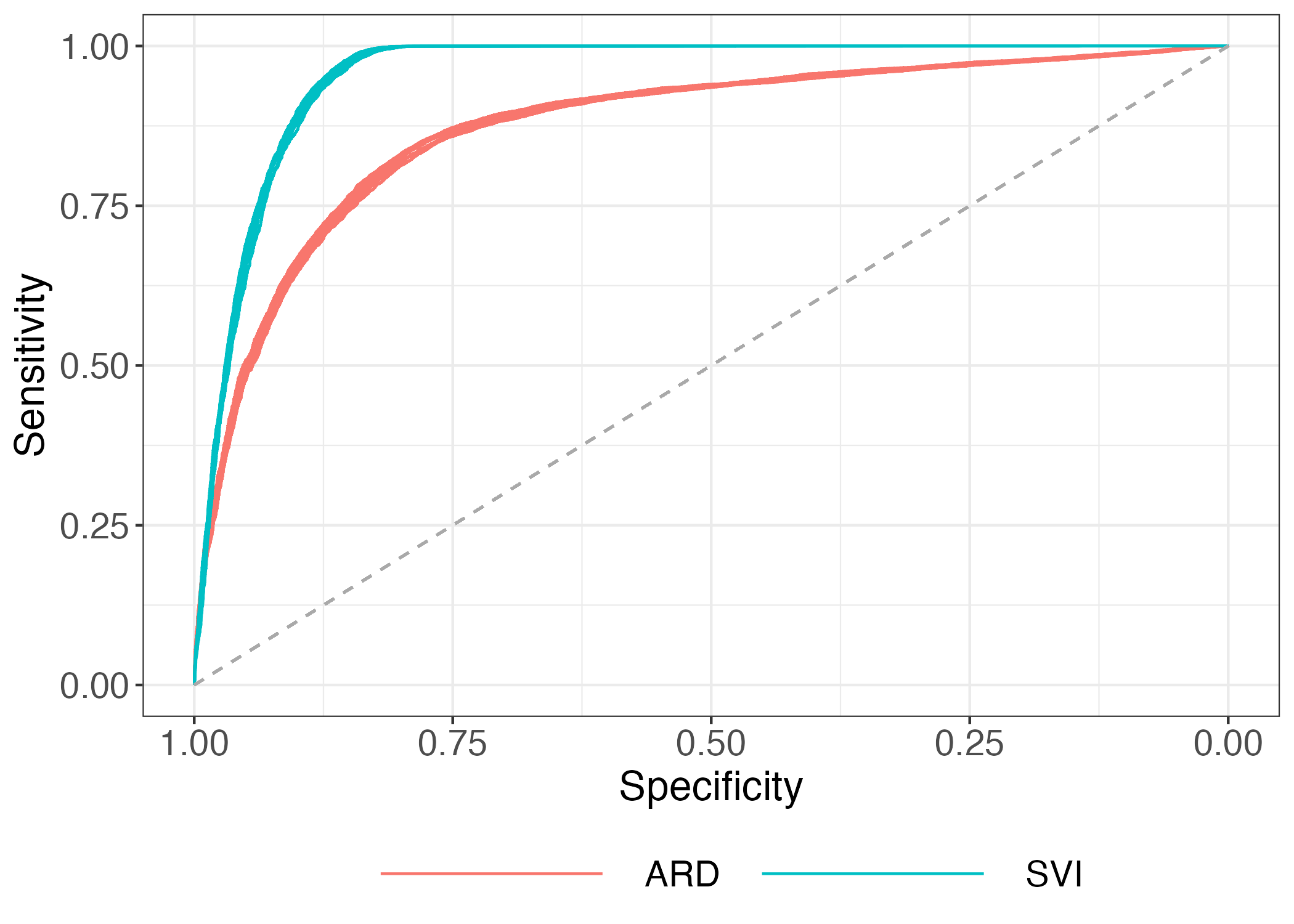} 
\end{minipage}
\begin{minipage}{0.25\textwidth}
\includegraphics[width=\linewidth]{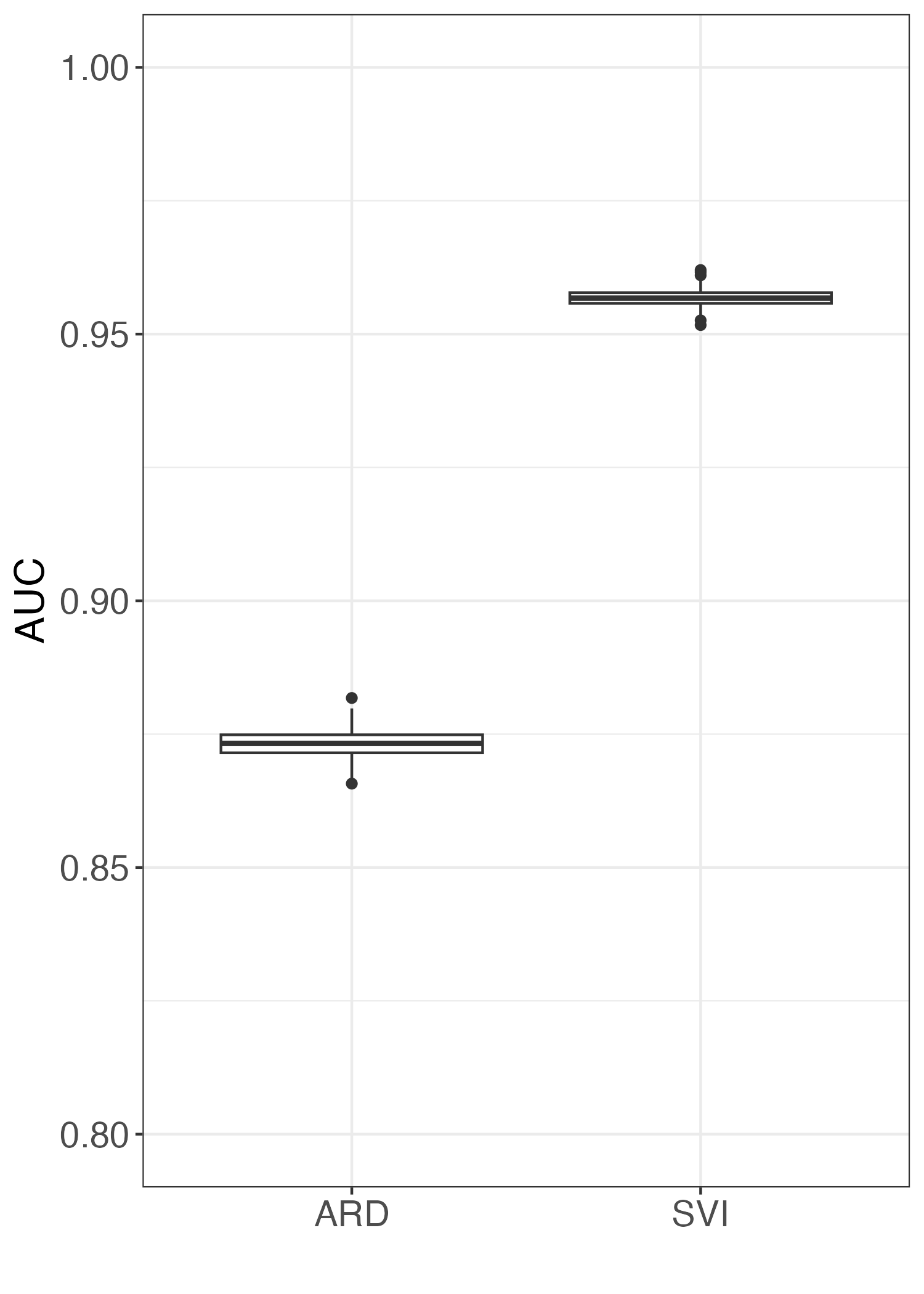}
\end{minipage}
\begin{minipage}{0.25\textwidth}%
\includegraphics[width=\linewidth]{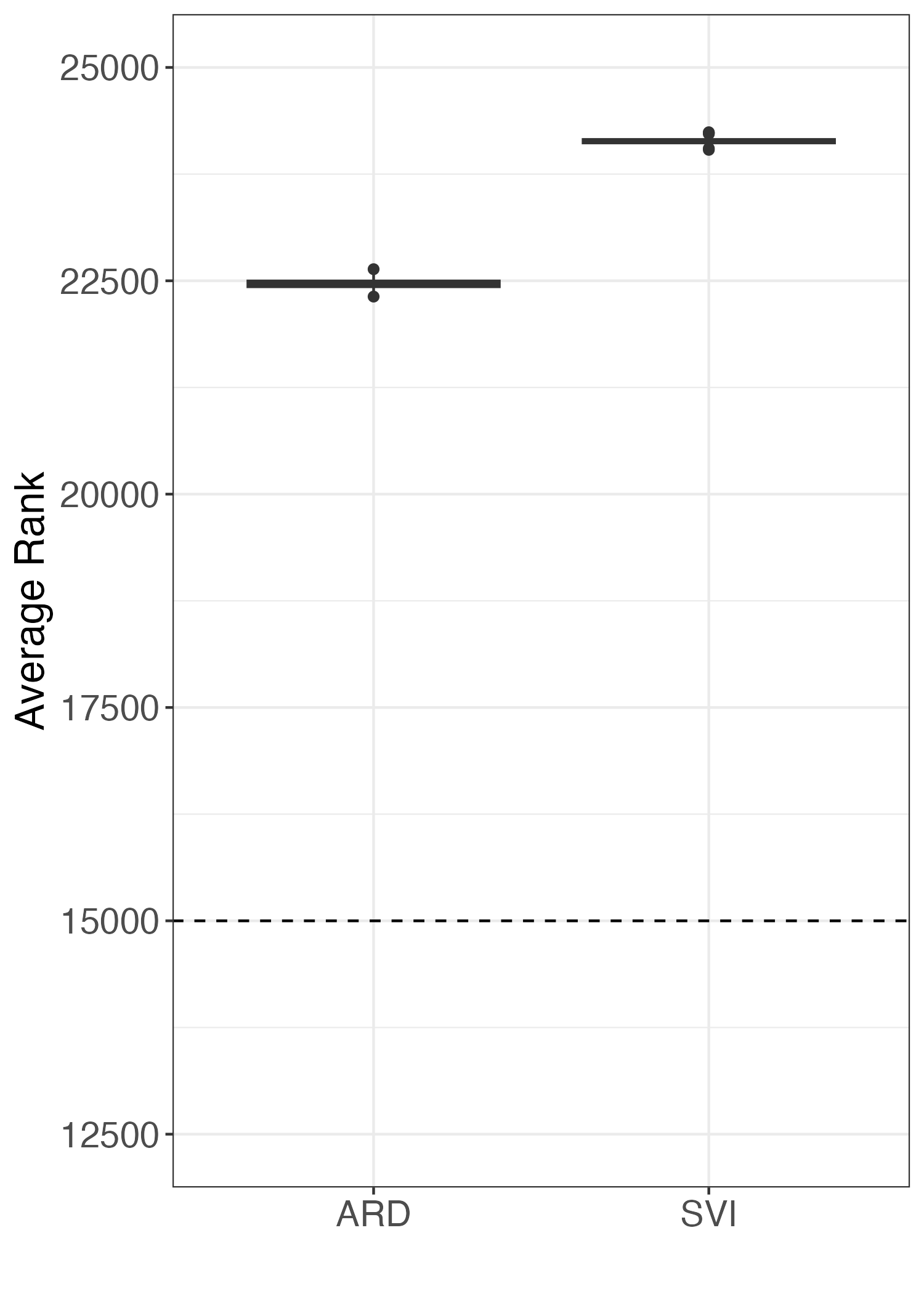}
\end{minipage}

\caption{Comparison between ARDMMSB and Gopolan \textit{et. al.}'s Stochastic Variational Inference. Left: ROC curves. Middle: Boxplot of the average area under the curve. Right: Boxplot of relative ranks.}
\label{fig:emp4}
\end{figure}

Figure~\ref{fig:emp4} shows quantitative measures of comparison between our inference procedure and \citet{gopalan2013efficient}'s stochastic variational inference algorithm. The left panel shows the ROC curves. The middle and right panels show boxplots of the area under the curve and relative ranks of both methods respectively. These plots were formed by subsampling 10,000 links and nonlinks and computing the probability of a link. Each of these subsamples forms an average rank measure and ROC curve which admits an area under the curve. The rank measure was formed by taking the average rank of the links according to their predictive probabilities. We see that our method performs worse with regards to these metrics. 

Although ARDMMSB does not do as well according to these measures, we only pick up communities formed by topics. ARD does not use paper to paper links. This can be beneficial as it removes some of the noise in link formation within the network. Since journals are human-curated by topics, aggregating links by journals ensure that ARD most likely uses link structure driven by topics. By only using paper to journal links, ARD removes some of the nuances due to other social structure such as community structure formed from coauthorships. \citet{gopalan2013efficient}'s algorithm may capture this kind of structure, which is different from structure solely from topics.
We have also created a Shiny app, available at \url{https://jyr123456.shinyapps.io/Topic_Terrain_Visualizer/}, to allow the exploration and interactive 
visualization of these results, allowing to user to specify a selection 
of topics and journals. 
\section{Conclusion}
\label{sec:conclusion}

The MMSB is popular model-based method of community detection of overlapping communities. However, this model is difficult to scale to networks on the order of millions of nodes since it parameterizes each pair of nodes. The algorithm from \citet{gopalan2013efficient} implements a stochastic variational inference algorithm with a sampling scheme that takes advantage of the sparseness found in real-world networks. Although this algorithm works quite well, it ignores nodal information commonly found in networks that could reduce computational costs and improve model convergence.

In this paper, we introduced aggregated relational data for the MMSB (ARDMMSB) that incorporates such information with the present link structure. ARDMMSB lends itself to a mini-batch strategy that can be carried out in parallel which can drive down computation times. It works by aggregating sets of links to form a bipartite graph which is of much smaller dimension than the network's adjacency matrix. It retains information from the original graph and can be used to estimate the parameters of the MMSB directly. Our simulation studies and real-world application to a citation network illustrate how our algorithm can achieve improved results by leveraging information outside of the links themselves. 
{\color{red}There are several important extensions of this procedure which should be considered in future work.
\citet{jin2023mixed} have extended the MMSB model to incorporate a degree correction parameter, 
which leads to better model fitting with real networks. Incorporating such a degree correction here 
is an important future direction.}
We have also made all code used to create the figures in this paper publicly available
in a github repo. 

\newpage
\appendix
\section{Appendix}
Here we provide the exact variational updates and
procedure which were ommitted from the main text.


\subsection{Variational Updates}

The approximate lower bound $\mathcal{L}^*$ is evaluated as 
\begin{align*}
\sum_{i,k,m,n} &y_{ik} \p \left[ \psi(\gamma_i^m) - \psi \left(\sum_d \gamma_i^d \right ) + \log B_{mn} + \psi(\phi_k^n) - \psi \left(\sum_d \phi_k^d \right ) \right] \\
& - \sum_{i,k,m,n} y_{ik} \p \log \p - \sum_{i,k,m,n} N_k \frac{\gamma_i^m}{\sum_d \gamma_i^d} B_{mn} \frac{\phi_k^n}{\sum_d \phi_k^d} \\
& + \sum_{i,m} (\alpha_m - 1) \left( \psi(\gamma_i^m) - \psi \left( \sum_d \gamma_i^d \right) \right) \\
& + \sum_{k,n} (\alpha_n - 1) \left( \psi(\phi_k^n) - \psi \left( \sum_d \phi_k^d \right) \right) \\
& - \sum_i \log \Gamma \left ( \sum_m \gamma_i^m \right ) + \sum_{i,m} \log \Gamma (\gamma_i^m) \\
& - \sum_{i,m} (\gamma_i^m - 1)\left( \psi(\gamma_i^m) - \psi \left( \sum_d \gamma_i^d \right) \right) \\
& - \sum_k \log \Gamma \left ( \sum_n \phi_k^n \right ) + \sum_{k,n} \log \Gamma (\phi_k^n) \\
& - \sum_{k,n} (\phi_k^n - 1)\left( \psi(\phi_k^n) - \psi \left( \sum_d \phi_k^d \right) \right) \\
& + \sum_{m,n} (a_{mn} - 1) \log B_{mn} +  \sum_{m,n} (b_{mn} - 1) \log (1 - B_{mn}).
\end{align*}

The variational distributions $q(\pi_i)$ and $q(\eta_k)$ belong to the exponential family. Thus we can derive updates by applying nonconjugate variational message passing. The updates for $\gamma_i$ and $\phi_k$ are

\[
\begin{bmatrix} 
\hat{\gamma}_i^1 - 1 \\
\hat{\gamma}_i^2 - 1 \\
\vdots \\
\hat{\gamma}_i^d - 1
\end{bmatrix} = \textit{I}^{-1}_{\gamma_i} \nabla_{\gamma_i} \expect_q [\log p(y, \Theta)], \quad 
\begin{bmatrix} 
\hat{\phi}_k^1 - 1 \\
\hat{\phi}_k^2 - 1 \\
\vdots \\
\hat{\phi}_k^d - 1
\end{bmatrix} = \textit{I}^{-1}_{\phi_k} \nabla_{\phi_k} \expect_q [\log p(y, \Theta)].
\]

$\textit{I}^{-1}_{\gamma_i}$ and $\textit{I}^{-1}_{\phi_k}$ are the inverse Fisher information matrices of Dirichlet$(\gamma_i)$ and Dirichlet$(\phi_k)$ respectively. The gradients are given by
\begin{align*}
\frac{\partial \expect_q \log p(y, \Theta)}{\partial \gamma_i^m} = \sum_{k,n} &y_{ik} p_{ik}^{(mn)} \left( \psi'(\gamma_i^m) - \psi' \left(\sum_d \gamma_i^d \right ) \right) \\
& - \frac{\sum_d \gamma_i^d - \gamma_i^m}{(\sum_d \gamma_i^d)^2} \sum_{k,n} N_k B_{mn} \frac{\phi_k^n}{\sum_d \phi_k^d} \\
& + \sum_{k,n,m' \neq m} N_k \frac{\gamma_i^{m'}}{(\sum_d \gamma_i^d)^2} B_{m'n} \frac{\phi_k^n}{\sum_d \phi_k^d} \\
& + (\alpha_m - 1)(\psi'(\gamma_i^m) - \psi'(\sum_d \gamma_i^d)) \\
& - \sum_{m' \neq m} (\alpha_{m'} - 1) \psi' ( \sum_d \gamma_i^d ). 
\end{align*}
and
\begin{align*}
\frac{\partial \expect_q \log p(y, \Theta)}{\partial \phi_k^n} = \sum_{i,m} &y_{ik} p_{ik}^{(mn)} \left( \psi'(\phi_k^n) - \psi' \left(\sum_d \phi_k^d \right ) \right) \\
& - \frac{\sum_d \phi_k^d - \phi_k^n}{(\sum_d \gamma_{G_k}^d)^2} \sum_{i,m} N_k \frac{\gamma_{i}^m}{\sum_d \gamma_i^d} B_{mn} \\ 
& + \sum_{i,m,n' \neq n} N_k \frac{\gamma_i^m}{\sum_d \gamma_i^d} B_{mn'} \frac{\phi_k^{n'}}{(\sum_d \phi_k^d)^2} \\
& + (\alpha_n - 1)(\psi'(\phi_k^n) - \psi'(\sum_d \phi_k^d)) \\
& - \sum_{n' \neq n} (\alpha_{n'} - 1) \psi' (\sum_d \phi_k^d). 
\end{align*}

The entries of the blockmatrix $B_{mn}$ are updated through a gradient descent given by

\begin{align*}
\frac{\partial \mathcal{L}^*}{\partial B_{mn}} = &B_{mn}^{-1} \left( \sum_{i,k} y_{ik}p_{ik}^{(mn)} + a_{mn} - 1 \right ) + (1 - B_{mn})^{-1}(1 - b_{mn}) \\
& - \sum_{i,k} N_k \frac{\gamma_i}{\sum_d \gamma_i^d} \frac{\phi_k^n}{\sum_d \phi_k^d}.
\end{align*}

Notice if we give the blockmatrix $B$ the noninformative prior Beta$(1,1)$, then we will have the closed from update
\[
\hat{B}_{mn} = \frac{\sum_{i,k} y_{ik}p_{ik}^{(mn)}}{\sum_{i,k} N_k \frac{\gamma_i^m}{\sum_d \gamma_i^d} \frac{\phi_k^n}{\sum_d \phi_k^d}}.
\]

The update for the ancillary parameters $p_{ik}^{(mn)}$ is based on tightening inequality 3.3. The tightest lower bound is given by
\[
p_{ik}^{(mn)} \propto \exp \left [ \expect_q \log(\pi_i^m B \eta_k^n) \right ].
\]

\subsection{Variational Inference Procedure for Multiple Passes}
Here we include the exact algorithm to combine the results obtained
from fitting ARDMMSB to multiple minibatches of the ARD network.

\begin{algorithm}[ht]
	\vspace{1mm}
	\small
	\begin{flushleft}
		Initialize {\color{red} variational parameters} $\gamma$, $\phi$, $B$.
	\end{flushleft}
	\begin{enumerate}
		
		\item Create minibatches by randomly partitioning nodes and using subsets of subpopulations. ({\color{red} Each subpopulation is expected to be present 
                    in multiple minibatches.})
		
		\item Apply each minibatch through steps 3 to 8 of Algorithm 1 in parallel.
		
		\item Collect outputs from each minibatch fit. Store the node parameters $\gamma_i$. Average the subpopulation parameters $\phi_k$ and blockmatrix $B$ across minibatches.
		
		\item Go back to step 1 until parameter estimates stabilize.
				
	\end{enumerate}
	\caption{Variational inference procedure for Multiple Passes}
	\label{Algorithm 3}
\end{algorithm}



\printbibliography

@article{fortunato2016community,
  title={Community detection in networks: A user guide},
  author={Fortunato, Santo and Hric, Darko},
  journal={Physics reports},
  volume={659},
  pages={1--44},
  year={2016},
  publisher={Elsevier}
}

@article{nowicki2001estimation,
  title={Estimation and prediction for stochastic blockstructures},
  author={Nowicki, Krzysztof and Snijders, Tom A B},
  journal={Journal of the American statistical association},
  volume={96},
  number={455},
  pages={1077--1087},
  year={2001},
  publisher={Taylor \& Francis}
}

@article{holland1983stochastic,
  title={Stochastic blockmodels: First steps},
  author={Holland, Paul W and Laskey, Kathryn Blackmond and Leinhardt, Samuel},
  journal={Social networks},
  volume={5},
  number={2},
  pages={109--137},
  year={1983},
  publisher={Elsevier}
}

@inproceedings{soundarajan2012using,
  title={Using community information to improve the precision of link prediction methods},
  author={Soundarajan, Sucheta and Hopcroft, John},
  booktitle={Proceedings of the 21st international conference on world wide web},
  pages={607--608},
  year={2012}
}

@article{ward2021next,
  title={Next waves in veridical network embedding},
  author={Ward, Owen G and Huang, Zhen and Davison, Andrew and Zheng, Tian},
  journal={Statistical Analysis and Data Mining: The ASA Data Science Journal},
  volume={14},
  number={1},
  pages={5--17},
  year={2021},
  publisher={Wiley Online Library}
}

@article{jin2023mixed,
  title={Mixed membership estimation for social networks},
  author={Jin, Jiashun and Ke, Zheng Tracy and Luo, Shengming},
  journal={Journal of Econometrics},
  year={2023},
  publisher={Elsevier},
  pages = {105369},
  issn = {0304-4076},
}

@article{geisser1975predictive,
  title={The predictive sample reuse method with applications},
  author={Geisser, Seymour},
  journal={Journal of the American statistical Association},
  volume={70},
  number={350},
  pages={320--328},
  year={1975},
  publisher={Taylor \& Francis Group}
}

@article{airoldi2008mixed,
	title={Mixed membership stochastic blockmodels},
	author={Airoldi, Edoardo M and Blei, David M and Fienberg, Stephen E and Xing, Eric P},
	journal={Journal of Machine Learning Research},
	volume={9},
	number={Sep},
	pages={1981--2014},
	year={2008}
}

@inproceedings{knowles2011non,
	title={Non-conjugate variational message passing for multinomial and binary regression},
	author={Knowles, David A and Minka, Tom},
	booktitle={Advances in Neural Information Processing Systems},
	pages={1701--1709},
	year={2011}
}

@article{mccormick2015latent,
	title={Latent surface models for networks using Aggregated Relational Data},
	author={McCormick, Tyler H and Zheng, Tian},
	journal={Journal of the American Statistical Association},
	volume={110},
	number={512},
	pages={1684--1695},
	year={2015},
	publisher={Taylor \& Francis}
}

@article{gopalan2013efficient,
	title={Efficient discovery of overlapping communities in massive networks},
	author={Gopalan, Prem K and Blei, David M},
	journal={Proceedings of the National Academy of Sciences},
	year={2013},
    volume={110},
    number={36},
    pages={14534--14539},
	publisher={National Acad Sciences}
}

@article{hoffman2013stochastic,
	title={Stochastic variational inference},
	author={Hoffman, Matthew D and Blei, David M and Wang, Chong and Paisley, John},
	journal={The Journal of Machine Learning Research},
	volume={14},
	number={1},
	pages={1303--1347},
	year={2013},
	publisher={JMLR. org}
}

@article{tan2016topic,
	title={Topic-adjusted visibility metric for scientific articles},
	author={Tan, Linda SL and Chan, Aik Hui and Zheng, Tian and others},
	journal={The Annals of Applied Statistics},
	volume={10},
	number={1},
	pages={1--31},
	year={2016},
	publisher={Institute of Mathematical Statistics}
}

@article{blei2017variational,
	title={Variational inference: A review for statisticians},
	author={Blei, David M and Kucukelbir, Alp and McAuliffe, Jon D},
	journal={Journal of the American Statistical Association},
	volume={112},
	number={518},
	pages={859--877},
	year={2017},
	publisher={Taylor \& Francis}
}

@article{bickel2009nonparametric,
	title={A nonparametric view of network models and Newman--Girvan and other modularities},
	author={Bickel, Peter J and Chen, Aiyou},
	journal={Proceedings of the National Academy of Sciences},
	pages={pnas--0907096106},
    volume={106},
    number={50},
	year={2009},
	publisher={National Acad Sciences}
}

@article{diprete2011segregation,
	title={Segregation in social networks based on acquaintanceship and trust},
	author={DiPrete, Thomas A and Gelman, Andrew and McCormick, Tyler and Teitler, Julien and Zheng, Tian},
	journal={American Journal of Sociology},
	volume={116},
	number={4},
	pages={1234--83},
	year={2011},
	publisher={University of Chicago Press Chicago, IL}
}

@article{mccarty2001comparing,
	title={Comparing two methods for estimating network size},
	author={McCarty, Christopher and Killworth, Peter D and Bernard, H Russell and Johnsen, Eugene C and Shelley, Gene A},
	journal={Human organization},
    volume={60},
    number={1},
    pages={28--39},
    year={2001},
	publisher={Society for Applied Anthropology}
}

@article{shelley1995knows,
	title={Who knows your HIV status? What HIV+ patients and their network members know about each other},
	author={Shelley, Gene A and Bernard, H Russell and Killworth, Peter and Johnsen, Eugene and McCarty, Christopher},
	journal={Social networks},
	volume={17},
	number={3-4},
	pages={189--217},
	year={1995},
	publisher={Elsevier}
}

@article{RN2,
   author = {McCormick, Tyler and Zheng, Tian},
   title = {Latent demographic profile estimation in hard-to-reach groups},
   journal = {The Annals of Applied Statistics},
   volume = {6},
   number = {4},
   pages = {1795-1813},
   ISSN = {1932-6157},
   DOI = {citeulike-article-id:12127567
doi: 10.1214/12-aoas569},
   url = {http://dx.doi.org/10.1214/12-aoas569},
   year = {2013},
   type = {Journal Article}
}

@article{RN4,
   author = {McCormick, Tyler H and He, Ran and Kolaczyk, Eric and Zheng, Tian},
   title = {Surveying hard-to-reach groups through sampled respondents in a social network},
   journal = {Statistics in Biosciences},
   volume = {4},
   number = {1},
   pages = {177-195},
   ISSN = {1867-1764},
   year = {2012},
   type = {Journal Article}
}

@article{RN5,
   author = {McCormick, Tyler H and Moussa, Amal and Ruf, Johannes and DiPrete, Thomas A and Gelman, Andrew and Teitler, Julien and Zheng, Tian},
   title = {A practical guide to measuring social structure using indirectly observed network data},
   journal = {Journal of Statistical Theory and Practice},
   volume = {7},
   number = {1},
   pages = {120-132},
   ISSN = {1559-8608},
   year = {2013},
   type = {Journal Article}
}

@inproceedings{leskovec2006sampling,
  title={Sampling from large graphs},
  author={Leskovec, Jure and Faloutsos, Christos},
  booktitle={Proceedings of the 12th ACM SIGKDD international conference on Knowledge discovery and data mining},
  pages={631--636},
  year={2006},
  organization={ACM}
}

@article{danon2005comparing,
  title={Comparing community structure identification},
  author={Danon, Leon and Diaz-Guilera, Albert and Duch, Jordi and Arenas, Alex},
  journal={Journal of Statistical Mechanics: Theory and Experiment},
  volume={2005},
  number={09},
  pages={P09008},
  year={2005},
  publisher={IOP Publishing}
}

@inproceedings{tang2008arnetminer,
  title={Arnetminer: extraction and mining of academic social networks},
  author={Tang, Jie and Zhang, Jing and Yao, Limin and Li, Juanzi and Zhang, Li and Su, Zhong},
  booktitle={Proceedings of the 14th ACM SIGKDD international conference on Knowledge discovery and data mining},
  pages={990--998},
  year={2008},
  organization={ACM}
}

@article{ma2017stabilized,
	title={Stabilized sparse online learning for sparse data},
	author={Ma, Yuting and Zheng, Tian},
	journal={The Journal of Machine Learning Research},
	volume={18},
	number={1},
	pages={4773--4808},
	year={2017},
	publisher={JMLR. org}
}

@article{zhao2012consistency,
	title={Consistency of community detection in networks under degree-corrected stochastic block models},
	author={Zhao, Yunpeng and Levina, Elizaveta and Zhu, Ji and others},
	journal={The Annals of Statistics},
	volume={40},
	number={4},
	pages={2266--2292},
	year={2012},
	publisher={Institute of Mathematical Statistics}
}

\supplementsection
\setcounter{subsection}{0}
\section*{Supplementary Materials}
{\color{red} \subsection{Additional Simulation Results}
Here we include additional experiments comparing the proposed ARDMMSB model with existing methods for simulated data.
While we have considered the performance of our proposed procedure in the main text, here we aim to more thoroughly
explore this, considering in turn several properties of both the data and the proposed inference scheme.
We also provide additional details on the simulation setting considered
in both the main paper and the simulation results provided here.

\paragraph{Computational Comparison}
As the focus of our proposed ARDMMSB method is to provide a scalable inference scheme for
large networks, we wish to examine the computational performance relative to 
the existing method for MMSM data of \citet{gopalan2013efficient}. One challenge here is that the
convergence metrics
differ for the two models. For the ARDMMSB model we can measure convergence through $\mathcal{L}^*$,
the approximate evidence lower bound. \citet{gopalan2013efficient} use an alternative 
approach
to measure convergence, computing perplexity on a held out set of node pairs.
As these are different convergence measures, we instead look at the performance of the two procedures when
they have both run for a number of iterations corresponding to observing an equivalent number of 
nodes. Namely, for the subgraphs used for each model, we iterate each model, updating 
the parameters until the total number of nodes (and the corresponding ARD or edge data)
``seen" by the respective algorithms is the same.
We consider the same simulation setting as the 
main text, and we further expand on the details of that setting here. 
Namely, we simulate 10 networks, each generated according to a MMSB model 
with $K=6$ communities and $N=10000$ nodes. The community probability matrix $B$
contains 2 values in the diagonal, with half of its entries corresponding to $0.04$
and half of the entries corresponding to $0.1$.
In the simulations in the main text we set all off diagonal entries to be $0$,
however, in all simulations which follow we instead consider
that all off diagonal entries in $B$ have the same small value $(0.005)$, making this
setting somewhat more challenging.
To generate ARD data we require known underlying subpopulations and we generate $\kappa=50$
subpopulation centers. We then generate the underlying membership vectors
as Dirichlet draws from the corresponding subpopulation center, before generating
the corresponding edges conditional on these membership vectors.
We run both ARD and SVI on subgraphs of size $n=500$ from these networks,
running each algorithm
until they have each observed data corresponding to $10000$ nodes. This is 
sufficient for model convergence for each of the respective metrics. 
The average computational time for the two procedures is shown in Figure~\ref{fig:compute-time}.
The SVI procedure takes approximately 50\% longer than ARD.
Running the algorithms for this period gives a
similar performance comparison (in terms of community and parameter recovery) as shown 
in the main text, with ARD able to obtain better performance in significantly less computation time.

\begin{figure}[ht]
    \centering
    \includegraphics[width = 0.65\textwidth]{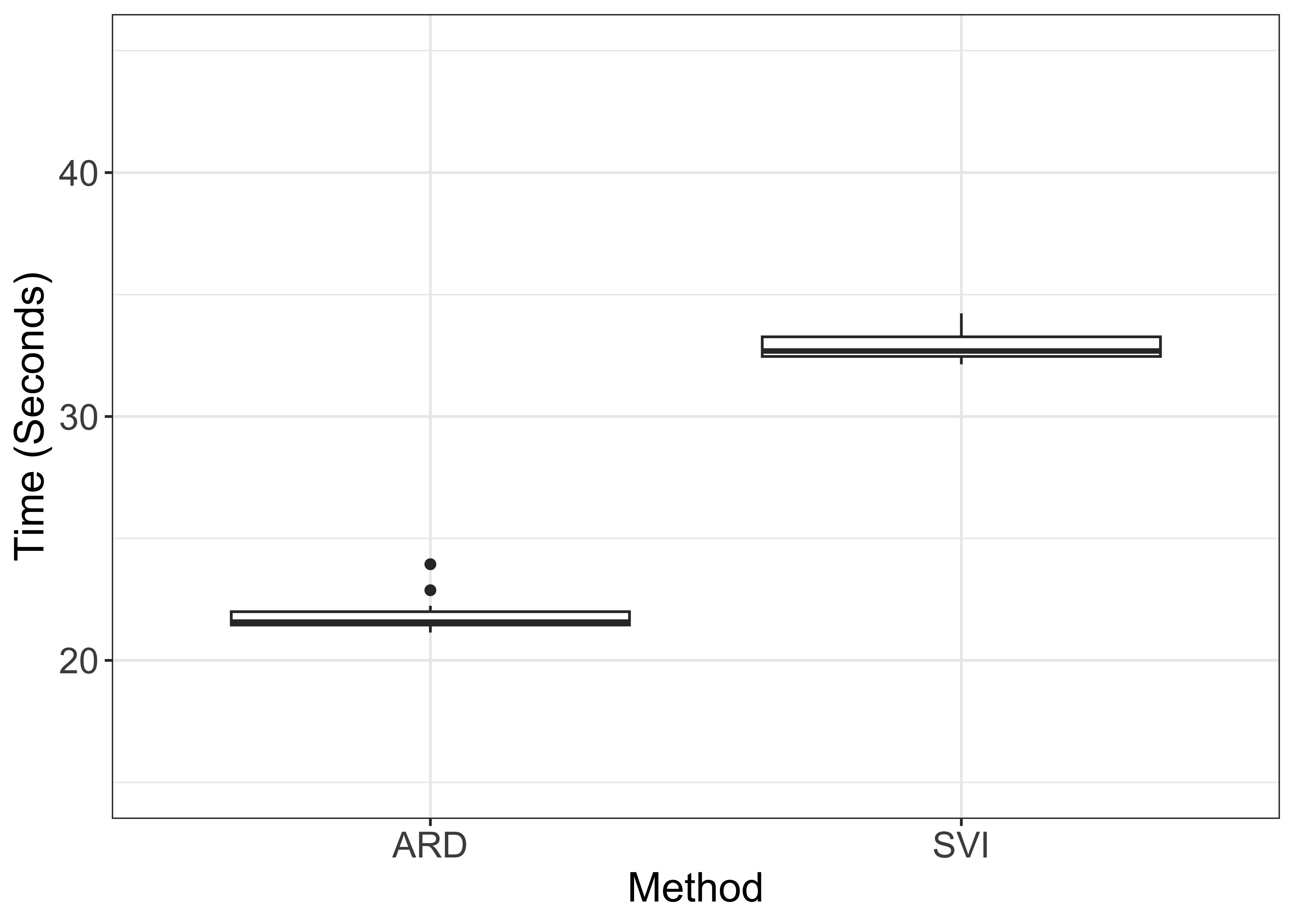}
    \caption{Boxplots of computation time for ARDMMSB and SVI, each observing the same number of events.}
    \label{fig:compute-time}
\end{figure}

\paragraph{Varying the Number of Communities}
To further examine the computational performance we consider the same setting 
described above, but we now vary the 
number of communities present in the data.
We consider $K=2,6,10,20$, in each case fitting ARD and SVI with
subgraphs of size $n=500$ along with SVI with the entire network of $n=10000$ nodes.
Community recovery and parameter recovery is shown in Figure~\ref{fig:vary-K}
When the number of communities is small both ARD and SVI can estimate the diagonal entries of $B$
well. As $K$ increases the performance of all procedures drops, with ARD showing clear improvement
over SVI with $n=500$ and also good performance against SVI with $n=10000$. In terms of
community recovery, ARD with subgraphs of size 500 and SVI using the entire network
show excellent community recovery, while SVI with a subgraph of size 500 cannot
recover the community structure, regardless of the number of communities present.


\begin{figure*}[h]
	\begin{minipage}{0.5\textwidth}
	    \centering
		\includegraphics[width=\textwidth]{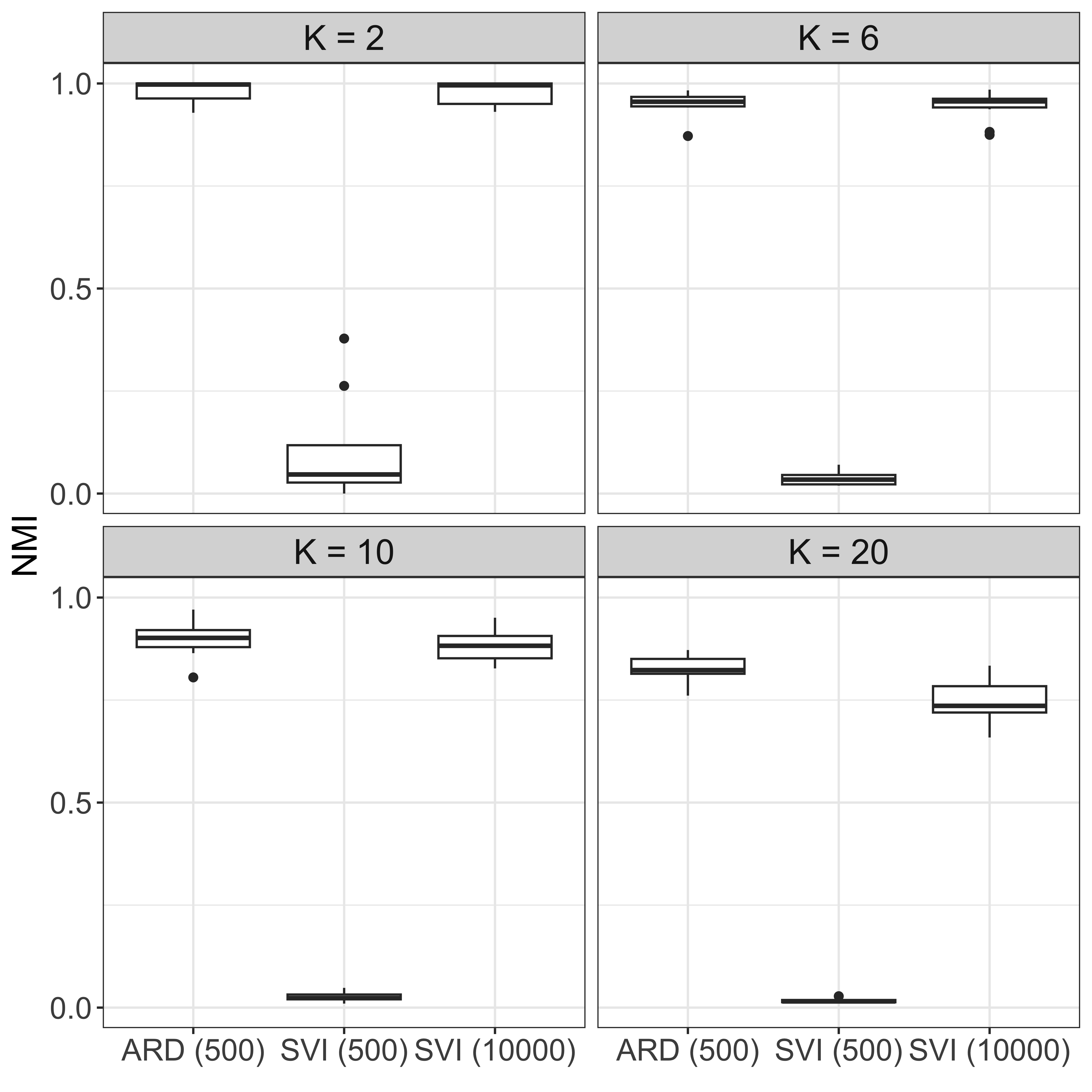}
	\end{minipage}
	\begin{minipage}{0.5\textwidth}
	    \centering
		\includegraphics[width=\textwidth]{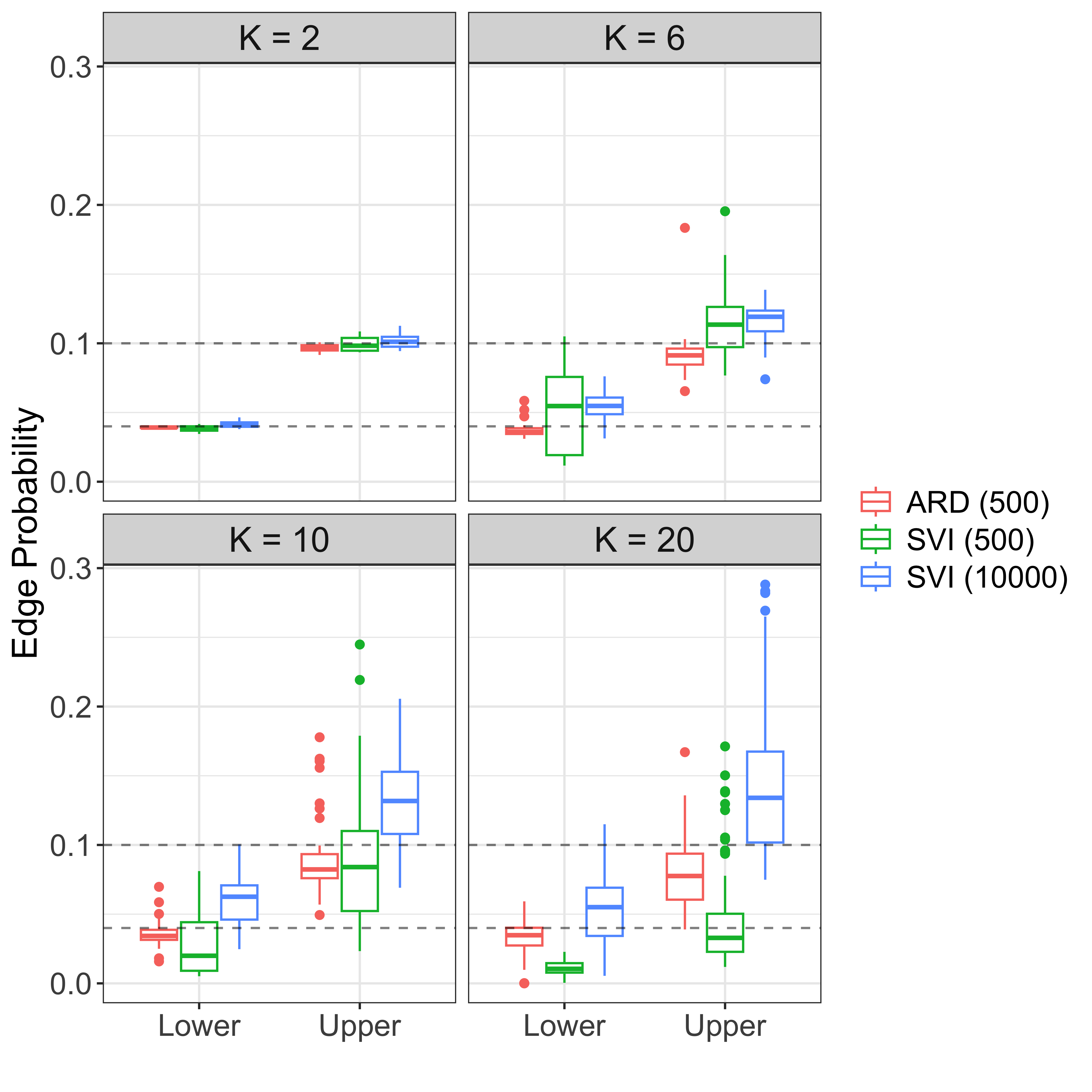}
	\end{minipage}
	\caption{Performance as we vary the number of communities, $K$. The left plot
            shows community recovery, in terms on NMI, while the right plots examines 
            recovery of the diagonal entries of $B$. In each case we consider ARD and SVI
            with subgraphs of $n=500$, along with SVI with the complete network.}
	\label{fig:vary-K}
\end{figure*}

\paragraph{Varying the Number of Subpopulations}
In the simulation settings in the main text we consider the number 
of subpopulations present in the network, $\kappa$,
to be fixed. Here we wish to examine how changing this number influences the overall results
and the performance of our proposed ARD approach.
We 
consider $\kappa = 10, 20,50,100$ and simulate data from the corresponding 
ARDMMSB model each time as above, with $K=6$ and all other parameters fixed as before.
In each case we again fit ARD and SVI
with subgraphs of size 500, along with applying SVI to the complete adjacency matrix.
Community and parameter recovery is shown in Figure~\ref{fig:vary-subpop}.
ARD with a subgraph of $n=500$ nodes performs as well as SVI with all nodes 
in terms of community recovery. As we increase $\kappa$, the uncertainty of the estimates
for the diagonal entries of $B$ decreases substantially. In these settings
the Poisson approximation in the ARDMMSB model is more appropriate,
indicating that the choice of many subpopulations, if possible, can lead to more stable
results.

\begin{figure*}[h]
	\begin{minipage}{0.5\textwidth}
	    \centering
		\includegraphics[width=\textwidth]{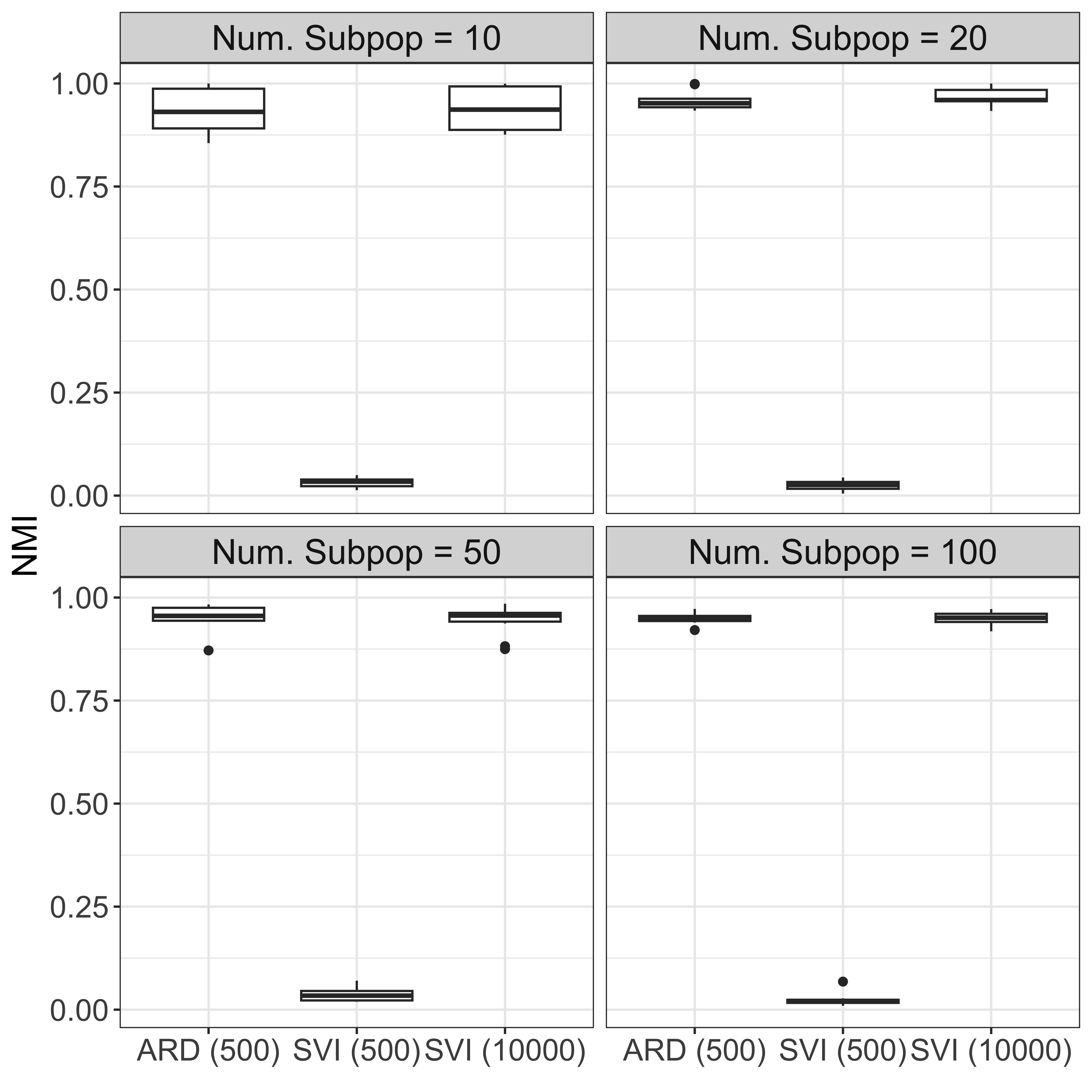}
	\end{minipage}
	\begin{minipage}{0.5\textwidth}
	    \centering
		\includegraphics[width=\textwidth]{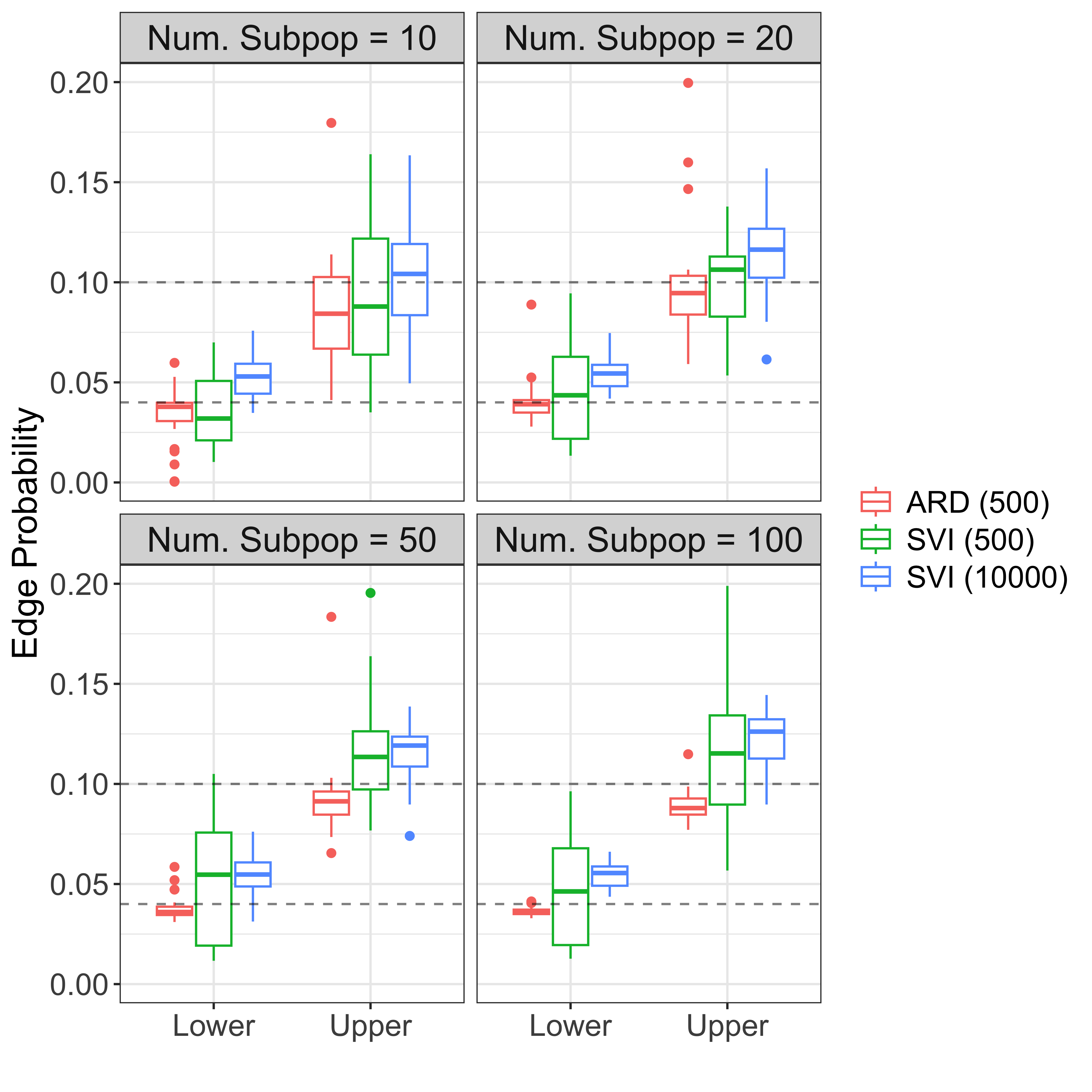}
	\end{minipage}
	\caption{Performance as the number of subpopulations in the network changes. The estimates
 of $B$ become more stable as the number of subpopulations increases.}
	\label{fig:vary-subpop}
\end{figure*}

\paragraph{Varying the Network Sparsity}
We also examine the role of network sparsity in the performance of our proposed inference
procedure. We consider three specific settings here. 
We consider the $B$ matrix described above,
with small constant off-diagonal value. The 
diagonal entries
consist of two values $B_{1},B_{2}$, with $K=6$ total communities and half of these communities
having connection probability $B_1$ and half having connection probability $B_2$. 
We examine the role of sparsity in these diagonal values, which drive the 
community structure in the network.
Utilising the constants $b_1=0.04$ and $b_2=0.1$, the sparsity settings we consider are:
\begin{itemize}
    \item Dense within community edge probability, matching the main simulations in the text, with $B_1=b_1$
    and $B_2=b_2$.
    \item Relatively sparse within community edge probability, where $B_1 =\frac{\log(N)}{N}b_1$ and 
    $B_2=\frac{\log(N)}{N}b_2$.
    \item Sparse within community edge probability, where $B_1=\frac{b_1}{N}$ and $B_2=\frac{b_2}{N}$.
\end{itemize}

Here $N$ is the number of nodes in the network, which is set to $N=10000$. We again examine community
and parameter recovery under each of these settings in Figure~\ref{fig:vary-spars}.
Unsurprisingly, when we consider sparse or relatively sparse connection 
probabilities, both ARD and SVI struggle. We do note that ARD with a subgraph of $n=500$
does seem to somewhat
outperform SVI (with subgraphs or the complete network), although all procedures
struggle.

\begin{figure*}
	\begin{minipage}{0.5\textwidth}
	    \centering
		\includegraphics[width=\textwidth]{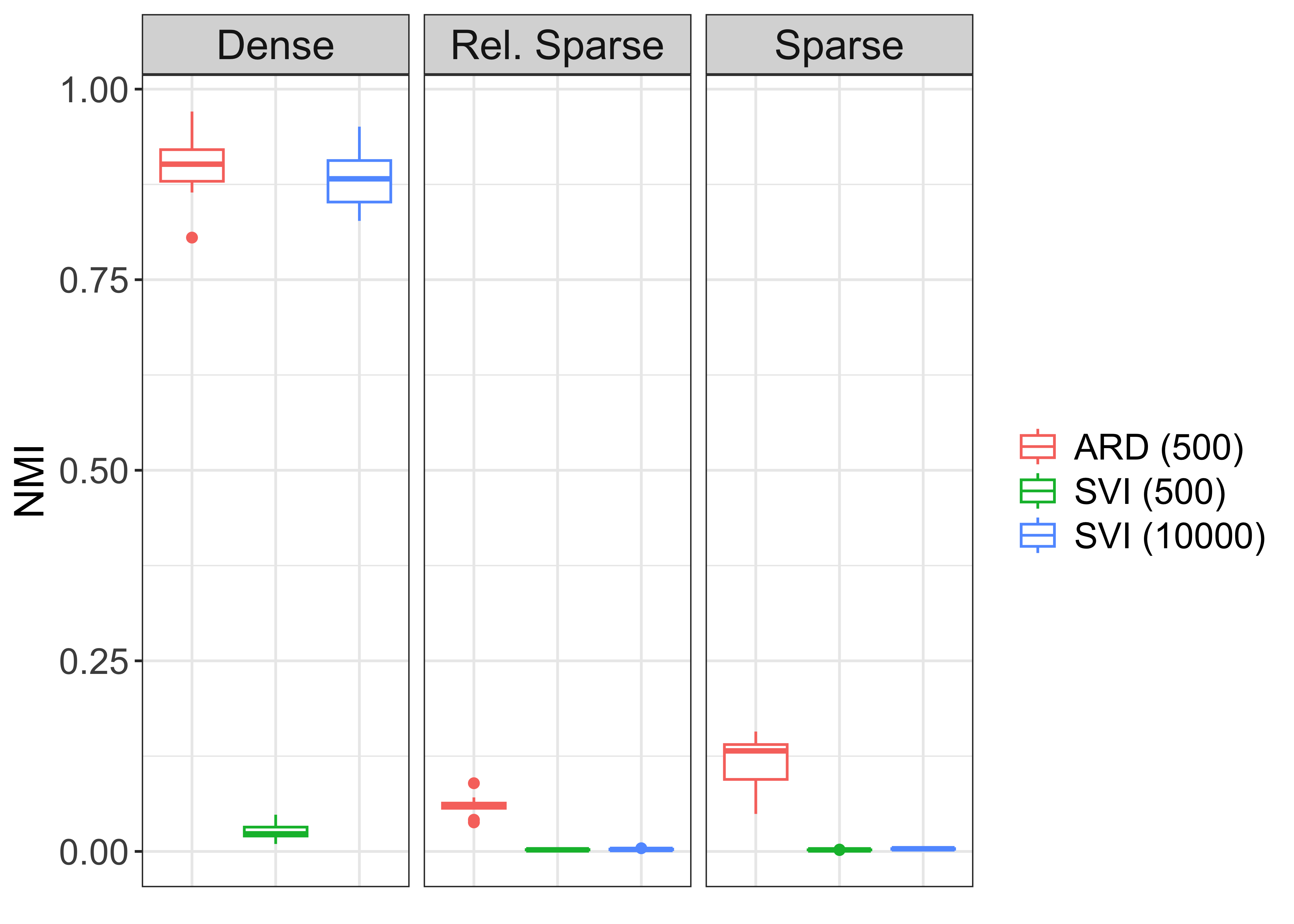}
	\end{minipage}
	\begin{minipage}{0.5\textwidth}
	    \centering
		\includegraphics[width=\textwidth]{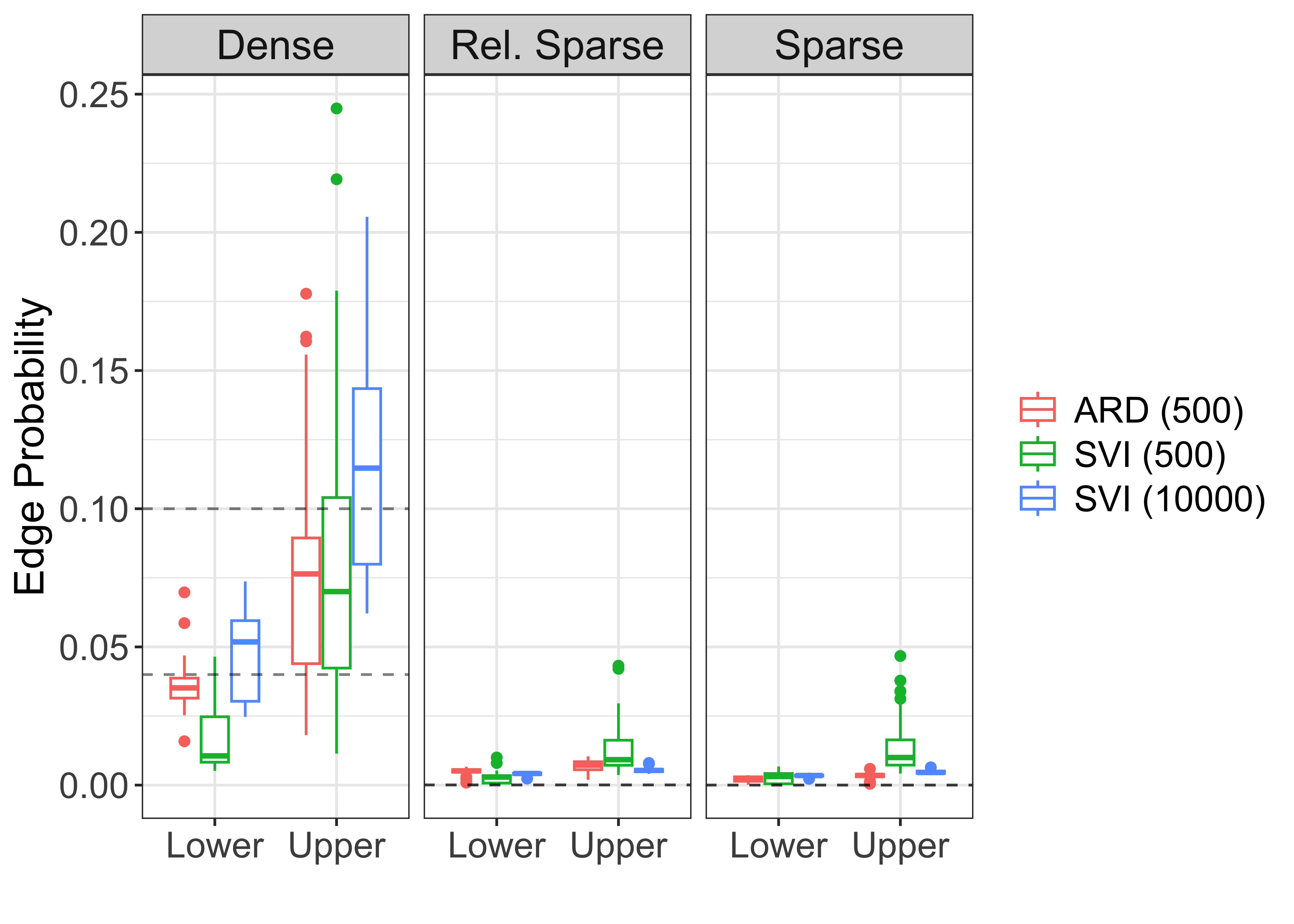}
	\end{minipage}
	\caption{Performance as the sparsity of the underlying network changes. 
            Performance decreases as the networks become sparser in all settings.}
	\label{fig:vary-spars}
\end{figure*}

}

\subsection{Multiple Passes for Citation Network Application.}
Our proposed algorithm fits the ARDMMSB model to minibatches in parallel. However, when fitting to a large network, each minibatch will contain a small fraction of nodes in the network. After initialization, the nodes in each minibatch will be run through the algorithm with weakly informative subpopulation blockmatrix parameters. Thus, the fit of each node ignores link information from all other minibatches. The subpopulation and blockmatrix parameters, on the other hand, will contain richer information since they are averaged over all the minibatches. We do another pass as summarized in Algorithm~S1 to allow this information to propagate back to the rest of the nodes.

\begin{figure}
    \centering
    \includegraphics[width = 0.8\textwidth]{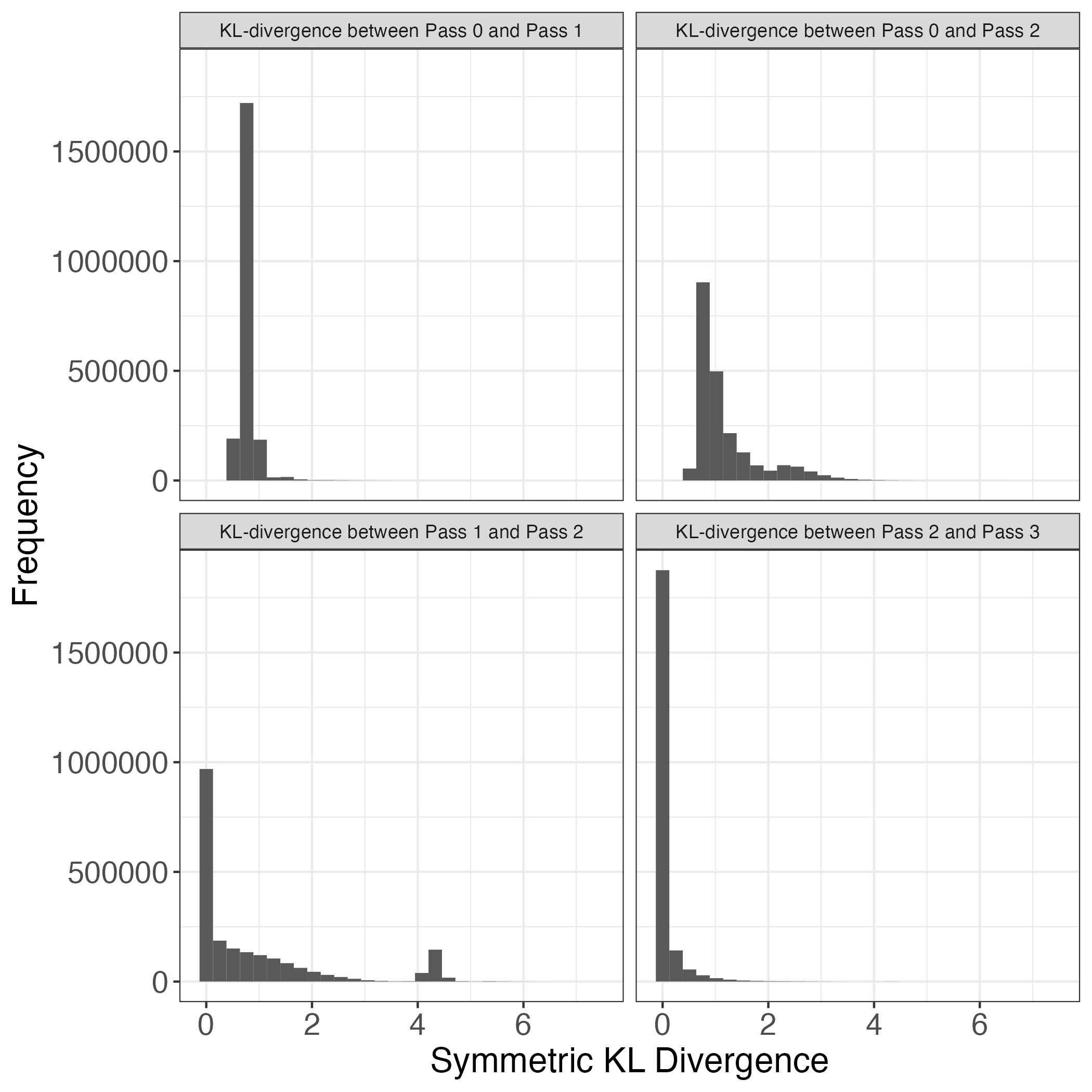}
    \caption{KL-Divergence between different passes of the community membership profile vectors of the papers in the Citation Network.}
    \label{fig:passes}
\end{figure}

Figure~\ref{fig:passes} illustrates the effect of running multiple passes on the membership profile vectors of the papers. Each plot is a histogram of the KL-divergences of the community membership profile vectors between passes of the algorithm. The top left plot shows that many of the papers did not move very far after the first pass. However, the top left plot shows that after the second pass, the paper membership profiles moved significantly. This illustrates that propagating the information that journals contain after the first pass is essential to update the paper profiles. After this propagation, another pass will not add much information and so the paper profiles will not change very much. This is clear in the bottom right plot.

\end{document}